\documentclass[conference]{IEEEtran}
\usepackage{amsmath}
\usepackage{amsthm}
\usepackage{amsfonts}
\usepackage{algorithm}
\usepackage{algorithmic}
\usepackage{graphicx}
\usepackage{multirow}
\usepackage{subfigure}
\usepackage[normalem]{ulem}

\useunder{\uline}{\ul}{}

\usepackage{array}
\newcolumntype{L}[1]{>{\raggedright\let\newline\\\arraybackslash\hspace{0pt}}m{#1}}
\newcolumntype{C}[1]{>{\centering\let\newline  \\\arraybackslash\hspace{0pt}}m{#1}}
\newcolumntype{R}[1]{>{\raggedleft\let\newline \\\arraybackslash\hspace{0pt}}m{#1}}

\graphicspath{ {img/} }

\def\bR{\mathbf{R}}
\def\bO{\mathbf{O}}
\def\bM{\mathbf{M}}

\def\bu{\mathbf{u}}
\def\bv{\mathbf{v}}

\def\bS{\mathbf{S}}

\def\bU{\mathbf{U}}
\def\bQ{\mathbf{Q}}
\def\bV{\mathbf{V}}

\newtheorem{theorem}{Theorem}[section]
\newtheorem{lemma}[theorem]{Lemma}
\theoremstyle{definition}

\begin{document}

\title{Collaborative Filtering with Social Local Models}

\author{\IEEEauthorblockN{Huan Zhao,  Quanming Yao$^{1}$, James T. Kwok and Dik Lun Lee}
	\IEEEauthorblockA{Department of Computer Science and Engineering \\
		Hong Kong University of Science and Technology \\
		Hong Kong\\
		\{hzhaoaf, qyaoaa, jamesk, dlee\}@cse.ust.hk}
}

\maketitle

\begin{abstract}
Matrix Factorization (MF) is a very popular method for recommendation systems. It assumes that the underneath rating matrix is low-rank. However, this assumption can be too restrictive to capture complex relationships and interactions among users and items. Recently, Local LOw-Rank Matrix Approximation (LLORMA) has been shown to be very successful in addressing this issue. It just assumes the rating matrix is composed of a number of low-rank submatrices constructed from subsets of similar users and items. Although LLORMA outperforms MF, how to construct such submatrices remains a big problem. Motivated by the availability of rich social connections in today's recommendation systems, we propose a novel framework, i.e., Social LOcal low-rank Matrix Approximation (SLOMA), to address this problem. To the best of our knowledge, SLOMA is the first work to incorporate social connections into the local low-rank framework. Furthermore, we enhance SLOMA by applying social regularization to submatrices factorization, denoted as SLOMA++. Therefore, the proposed model can benefit from both social recommendation and the local low-rank assumption. Experimental results from two real-world datasets, Yelp and Douban, demonstrate the superiority of the proposed models over LLORMA and MF.
\footnote{Quanming Yao is the Corresponding author.}
\end{abstract}

\vspace{5px}



\section{Introduction}
\label{sec-intro}
Recommendation System (RS) has become an indispensable tool in the big data era. It tackles information overload by helping users to get interesting items based on their previous behaviors.
Collaborative Filtering (CF), a state-of-the-art RS technique, tries to predict users' ratings (or preferences) on unseen items based on similar users or items.
Among various CF-based methods, Matrix Factorization (MF) is most popular due to its good performance and scalability~\cite{paterek2007improving,mnih2007probabilistic,koren2008factorization,yao2015accelerated}. MF is based on the assumption that users' preferences to items are controlled by a small number of latent factors. Thus, the large user-item rating matrix can be decomposed into two smaller matrices, representing user-specific and item-specific latent factors, respectively. In other words, the rating matrix is of \textit{low-rank}.

\begin{figure*}
\centering
\includegraphics[width=0.7\textwidth]{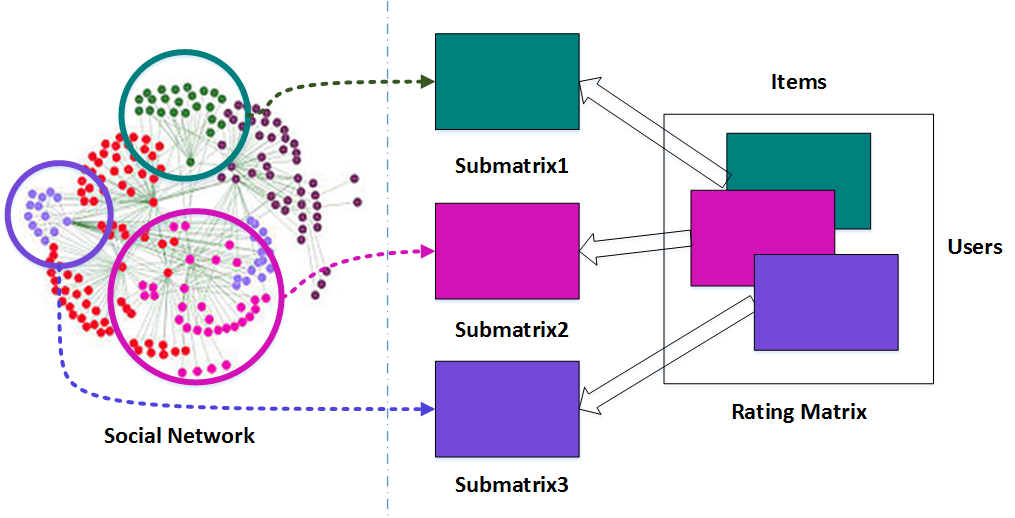}
\caption{An illustration of SLOMA. \textbf{Left-hand side:} an example social
network, where underlying social groups are circled. 
\textbf{Right-hand side:} the core idea of LLORMA, that is, the rating matrix is composed of a number of submatrices which are low-rank. SLOMA enhances LLORMA with the social groups underlying the social network to construct meaningful submatrices. Therefore, SLOMA can enjoy the advantages of both social recommendation and LLORMA, leading to better recommendation performance.}
\label{fig-example}
\end{figure*}

Despite the success of MF in RS, the assumption that the rating matrix is low-rank (termed \textit{global low-rank}) is problematic because the rating matrix in a real-world scenario is very large and composed of diverse rating behaviors, making the low-rank assumption of the rating matrix improper.
Lee et al.~\cite{lee2016llorma} proposed a novel framework called Local LOw-Rank Matrix Approximation (LLORMA) to alleviate the global low-rank problem.
Based on the observation that there tend to be groups of users who are interested in small sets of items,
LLORMA takes the view that the rating matrix is composed of a number of low-rank submatrices (termed \textit{local low-rank}), which is illustrated in the right part of Figure~\ref{fig-example}. Extensive experiments have demonstrated the effectiveness of LLORMA in recommendation systems~\cite{lee2016llorma,chen2015wemarec,chen2016mpma,zhang2017local}. Besides RS, subsequent works have been performed in different domains based on the local low-rank assumption, e.g., image processing~\cite{qyao2015aaai,li2016locality}, multi-label classifications~\cite{bhatia2015sparse}, document analysis~\cite{lee2016icdm}, demonstrating the efficacy of this framework.

In the local low-rank approach, the construction of the submatrices is a fundamental problem. Lee et~al. \cite{lee2016llorma} proposed to firstly choose some random anchor points, i.e., user-item pairs, from the rating matrix. Then, for each anchor point, one submatrix is constructed by selecting the remaining points that are close to the anchor point based on some distance metrics. However, this method leads to several problems. Firstly, since anchor points are randomly selected, it is hard to
explain the meaning of the obtained submatrices, and thus the recommendation results are also unexplainable. Secondly, since the submatrices are constructed around the anchor points using a distance threshold,
they are very sensitive to the distance threshold. Thus, finding a good distance threshold value is a challenging task. Finally, the computation and space costs are both high, because we need to compute and store the pair-wise similarities of users and items in all submatrices.

To overcome the weaknesses of LLORMA, we propose a novel framework, Social LOcal low-rank Matrix Approximation (SLOMA), which incorporates the social connections among users into the local low-rank framework. While social connections have been exploited effectively in MF-based RSs~\cite{ma2008sorec,jamali2010matrix,ma2011recommender,tang2013exploiting,guo2015trustsvd}, to the best of our knowledge, SLOMA is the first work to utilize social connections under the local low-rank framework. As in most social network research, SLOMA assumes that the social graph embeds a number of social groups, within which users have similar preferences to and influences on each other. The basic idea of SLOMA is illustrated in the left part of Figure~\ref{fig-example}. There are three social groups underlying the social network,\footnote{Note that the groups can overlap in reality, here we put them separately for simplicity.} based on which we can construct three submatrices out of the rating matrix satisfying the low-rank assumption. The social connections help SLOMA in solving the important submatrix construction problem. Specifically, instead of randomly selecting anchor points as in LLORMA, SLOMA can select influential users, termed \textit{connectors}, and pick the connectors' friends within a certain distance in the social graph (i.e., within several hops) to construct the submatrices, which are meaningful because they can be treated as social groups. See the example in Figure~\ref{fig-example}. Thus, SLOMA solves all the three problems faced by LLORMA, that is, it selects meaningful anchors and build meaningful submatrices, its distance measure is intuitive, and there is no need to keep the pairwise similarities of users and items for each submatrix, which avoids high computation and space costs.

In summary, the contributions of our work are as follows.
\begin{itemize}
\item 
To the best of our knowledge, SLOMA is the first work that
incorporates social connections into the local low-rank framework, which can enjoy the advantages from both sides, i.e., social recommendation and local low-rank framework.

\item 
SLOMA addresses the submatrice construction problem in LLORMA by exploiting social connections.
These submatrices have better interpretability (explained by social homophily theory)
and lead to superior prediction accuracy over LLORMA.

\item In addition to SLOMA, we also propose SLOMA++ that incorporates social regularization techniques into each local model to further improve the performance.

\item We conduct extensive experiments on two real-world datasets to show the effectiveness of our proposed methods. We not only demonstrate the effectiveness of SLOMA and SLOMA++, but also give insights into social recommendation under the local low-rank framework.
\end{itemize}

The rest of the paper is organized as follows. MF, LLORMA and social recommendation are introduced in Section \ref{sec-related}. We then elaborate our SLOMA model in Section \ref{sec-framework}, and the experiments as well as the analysis in Section \ref{sec-exp} and \ref{sec-analysis}, respectively. Finally, we conclude our work in Section \ref{sec-conclusion}.

\section{Background}
\label{sec-related}
\subsection{MF and LLORMA}
MF has been one of the most popular approaches for rating prediction in recommendation systems. Based on the assumption that users' preferences to items are governed by a small number of latent factors, the rating matrix $\bO \in \mathbb{R}^{m \times n}$ can be approximated by the product of $l$-rank matrices,
\begin{equation}
\label{eq-mf-predict}
\bR \approx \bU\bV^{\top},
\end{equation}
where $\bU \in \mathbb{R}^{m \times l}$ and {$\bV \in \mathbb{R}^{n \times l}$} with $l \ll min(m, n)$, representing,
respectively, the latent features of users' preferences and items. The approximating process can be completed by solving the following optimization problem:
\begin{equation}
\label{eq-mf-obj}
\min_{\bU,\bV}\frac{1}{2}\sum_{(i,j) \in \Omega}(\bO_{ij} - \bu_i\bv_j^{\top})^2,
\end{equation}
where $\bu_i$ and $\bv_j$ are the $i$th and $j$th rows of $\bU$ and $\bV$, representing the latent feature vectors of user $u_i$ and item $v_j$, respectively. $||\cdot||_F$ denotes the Frobenius norm, and $\Omega$ is the index set of the observations, and $\bO_{ij}$ is the observed rating of user $u_i$ to item $v_j$. In order to avoid overfitting, two regularization terms, $\frac{\lambda}{2}\left( ||\bU||_F^2 + ||\bV||_F^2 \right)$, are added to Equation~\eqref{eq-mf-obj}. In the literature, this method is also termed Regularized SVD (RegSVD)~\cite{paterek2007improving}.

Instead of assuming the rating matrix to be low-rank, Lee et al. proposed a novel framework LLORMA that assumes the rating matrix is \textit{local low-rank} \cite{lee2016llorma}, i.e., it is composed of a number of submatrices which are of low-rank. After obtaining the submatrices, MF is applied to each submatrix independently, and then a weighted ensemble scheme is designed to approximate the rating matrix. The key component of LLORMA is the construction of submatrices, which is depicted in Algorithm \ref{alg-llorma-sub-con}.

\begin{algorithm}[ht]
	\caption{Submatrice Construction (LLORMA)}
	\begin{algorithmic}[1]
		\REQUIRE{Observed index set $\Omega$, number of submatrices $q$, distance threshold $d_1,d_2$;}
		\FOR{$t=1,2,...,q$}
		\STATE{Select a point $(i_t,j_t)$ randomly from $\Omega$;}\label{alg-sc-anchor-select}
		\STATE{Obtain a subset of users $U^t = \big\{u_k:d(u_{i_t},u_k) \le d_1\big\}$;}\label{alg-sc-users-select}
		\STATE{Obtain a subset of items $V^t = \big\{v_k:d(v_{j_t},v_k) \le d_2\big\}$;}\label{alg-sc-items-select}
		\STATE{Obtain $\bM^t$; // \textit{The $t$th submatrix based on $U^t$ and $V^t$.} \label{alg-sc-submatrix-obtain}}
		\ENDFOR
		\RETURN{$\big\{\bM^t\big\}_{t=1}^q$}
	\end{algorithmic}
	\label{alg-llorma-sub-con}
\end{algorithm}

In Line~\ref{alg-sc-anchor-select}, a point ($i_t,j_t$), termed \textit{anchor point}, 
is chosen randomly.\footnote{There are more complex methods for picking up anchor points, 
but random choosing gives the best predicting performance as reported in~\cite{lee2016llorma}. Thus, we focus on random choosing here.} $d(\cdot)$ is a distance function for any two users or items in the rating matrix. $d_1$ and $d_2$ are two distance thresholds for selecting users or items according to Line \ref{alg-sc-users-select} and \ref{alg-sc-items-select}. A submatrix is then constructed from the selected users and items that are within the distance threshold from the anchor point. In LLORMA, the distance $d$ is based on the cosine similarity between any two latent features of users or items. 
It is reported that such measurement gives the best predicting performance \cite{lee2016llorma}.
Specifically, the arc-cosine between users $u_{i_t}$ and $u_k$ is
\begin{align}
d(i_t,k) = \arccos\bigg(\frac{\bu_{i_t}\bu_k}{||\bu_{i_t}||\cdot||\bu_k||}\bigg).
\label{eq-arccos}
\end{align}
Besides, the similarity between items are also computed based on arc-cosine distance at Equation~\eqref{eq-arccos}.

After obtaining $q$ submatrices, MF is applied to each submatrice independently, leading to $q$ groups of user-specific and item-specific latent features. Finally, the rating matrix $\bR$ is approximated by the weighted ensemble of product of these $q$ groups of user-specific and item-specific latent feature vectors as follows
\begin{equation}
\bR_{ij} = \sum_{t=1}^{q}\frac{w^t_{ij}}{\sum_{s=1}^{q}w^s_{ij}}\big[\bu^t_i(\bv^t_j)^{\top}\big],
\label{eq-llorma-ensemble}
\end{equation}
where $w^t_{ij}$ is the weight computed based on the distance between a point $(u_i,v_j)$
and an anchor point $(u_{i_t},v_{j_t})$ in every submatrix. $\bu^t_i$ and $\bv^t_j$
represent the latent feature vectors of $u_i$ and $v_j$, respectively, from the $t$th
submatrix. It means that the prediction of a point $(u_i,v_j)$ in the rating matrix is a
weighted average of the predicted ratings from all submatrices where $u_i$ and $v_j$ both
occur. Experiments on real-world datasets show that LLORMA improves prediction accuracy
compared to RegSVD. Besides
recommendation~\cite{lee2016llorma,chen2015wemarec,chen2016mpma,zhang2017local}, the local
low-rank assumption has been demonstrated to be effective in many other domains, including
image processing~\cite{qyao2015aaai,li2016locality}, multi-label
classifications~\cite{bhatia2015sparse}, and document analysis~\cite{lee2016icdm}.

\subsection{Social Recommendation in MF}
Social recommendation has become more and more popular with the proliferation of online social networks such as Facebook, Twitter and Yelp. Previous works have been performed to incorporate social connections into the MF framework
\cite{ma2008sorec,jamali2010matrix,ma2011recommender,guo2015trustsvd,li2015overlapping}. Most of these works are based
on the social homophily theory \cite{mcpherson2001birds} that people with similar preferences tend to be connected as friends. Thus, the latent feature vectors of friends should be closer after factorizing the rating matrix. One typical method is to add the social regularization term to the objective function~\eqref{eq-mf-obj}, which is defined in the following:
\begin{equation}
\frac{\beta}{2} \sum_{i=1}^{m}\sum_{j \in \mathcal{F}(i)}\bS_{ij}||\bu_i - \bu_j||_2^2,
\label{eq-soc-reg}
\end{equation}
where $\beta > 0$ is the weight of social regularization term, $\mathcal{F}(i)$
is the set of $u_i$'s friends, and $\bS_{ij}$ is the similarity between $u_i$ and $u_f$.
The more similar two users are, the closer their latent feature vectors.

We note that there are no previous work on social recommendation under the local low-rank
framework. The major difference between our SLOMA and LLORMA is that LLORMA uses random
anchor points to construct submatrices, while we use social connections. Thus, we can obtain
meaningful submatrices and better prediction ability because each submatrix can represent
the behavior of a socially connected user group. Moreover, we enhance SLOMA to SLOMA++ by
incorporating social regularization into every local model, and demonstrate with experiments that recommendation performance can be further improved.

\section{Social Local Matrix Approximation}
\label{sec-framework}

In this section, we first describe the main problem facing submatrix construction in LLORMA,
which also motivates our work. Then, we elaborate the proposed SLOMA framework.

\subsection{Motivation}
\label{subsec-frmamewok-motivation}

Despite the success of LLORMA, the main weakness of LLORMA is its submatrix construction method (Algorithm~\ref{alg-llorma-sub-con}). 
We identify three problems below.

First, the anchor points are chosen randomly, 
which do not have any reasonable justification and cannot be interpreted. 
Second, the constructed submatrices are neither stable nor meaningful.
Specifically, the cosine similarity in Equation~(\ref{eq-arccos}) is used in LLROMA,
which is computed in Lines~\ref{alg-sc-users-select} and \ref{alg-sc-items-select} of Algorithm~\ref{alg-llorma-sub-con}.
From Lemma~\ref{lemma-invert} (proof can be found in Appendix~\ref{lem:proof}), 
we can see that different pairs of ($\bU$, $\bV$) may be obtained from solving the optimization problem in Equation~(\ref{eq-mf-obj}).
However, the cosine similarity is not robust to the transformation in Lemma~\ref{lemma-invert}. Using the same anchor point with the pair ($\bU$, $\bV$) or ($\hat{\bU}$, $\hat{\bV}$), different submatrices can be constructed. 
As a result, the constructed submatrices are not consistent and then become meaningless. 
Third, to predict a missing rating in the original big rating matrix,
a weighted average scheme is used to combine the ratings in the submatrices according to Equation~\eqref{eq-llorma-ensemble}. This suffers from the same problem facing submatrix construction,
as the weights are also computed based on cosine similarity.

\begin{lemma} 
Given any matrices $\bU$ and $\bV$ which are an optimal solution to \eqref{eq-mf-obj},
then $\hat{\bU} = \bU \bQ$ and $\hat{\bV} = \bV \bQ^{-1}$ are also an optimal solution, 
if matrix $\bQ$ is invertible.
\label{lemma-invert}
\end{lemma}

Motivated by these problems, we propose a novel framework integrating social connections with the local low-rank framework. Based on the social homophily theory~\cite{mcpherson2001birds} that people with similar preferences tend to be connected as friends, we use the social connections among users as an explicit indicator of similarity between users' preferences in constructing the submatrices. Since each submatrix contains socially connected users with similar preferences, it satisfies the low-rank property. Moreover, we argue and demonstrate with experiments that the submatrices obtained in our model are better than those in LLORMA in terms of overall prediction ability. The core idea of our proposed model is illustrated in Figure~\ref{fig-example}.

\subsection{SLOMA}

In this section, we give a formal and mathematical elaboration of SLOMA and design the optimization approaches.

We first introduce the notations in this work. Let $\mathcal{U} =
\{u_1,u_2,...,u_m\}$ and $\mathcal{V} = \{v_1,v_2,...,v_n\}$ be the sets
of users and items, respectively. Let
$\mathcal{G} = (\mathcal{U},\mathcal{E})$ represents a social network
graph, where the vertex set $\mathcal{U} = \{u_1,u_2,...,u_m\}$ represents
the users and the edge set $\mathcal{E} = \{e_1,e_2,...,e_p\}$ represents
the social connections among all users. The weight of every edge of
$\mathcal{G}$ is set to 1, representing the existence of friendship
between two users.\footnote{We assume that the social graph is undirected
and unweighted.} Then, for any two users $u_i$, $u_j$, $d(u_i,u_j)$ is the
distance between $u_i$ and $u_j$, which can be computed from the social
graph. In this paper, we propose a novel framework to integrate social
recommendation with the local low-rank assumption. The key challenge is
how to construct submatrices from the rating matrix. We develop several
approaches, including heuristic and systematic approaches.

After constructing the submatrices, we apply MF to them and obtain a number of user-specific and item-specific latent features in different submatrices. It is obvious that these submatrices have overlaps, i.e., a user and an item can occur in more than one submatrices, which aligns with our intuition that users tend to participate in multiple groups in social networks. 

Overall, SLOMA consists of the following steps:
\begin{itemize}
	\item Identify $q$ social groups from the social graph $\mathcal{G}$,
	and then construct $q$ submatrices from the rating matrix $\bO$ based on those groups.
	\item Apply MF to all the submatrices, i.e., $\{\bM^t\}_{t=1}^q$, independently,
	and obtain $q$ groups of user-specific and item-specific latent features.
	\item Predict the missing ratings in $\bR$ using the ensemble of predictions from all submatrices.
\end{itemize}

\subsection{Construction of Submatrices}
\label{framework-sc}
In SLOMA, we assume that there are groups of users underlying the social network, whose preferences are similar due to the fact that they are socially connected. To identify these groups, we develop both heuristic and systematic approaches in this work.

\noindent
\textbf{Heuristic Approaches.}
For the heuristic approaches, we build the social groups based on the fact that users' influences to each other can propagate through the networks. We observe that users in a social group can affect each other and the amount of influence can vary. Intuitively, those with more friends tend to have larger influence than others. 
Based on this,
we construct the submatrices by first selecting a number of influential users, i.e., the connectors, and their friends within a fixed number of
hops, 
e.g., three, 
in the social network. 
Then, for each group of users, we select the items they rate and then the submatrix is constructed from this user-item group.
Note that there are several methods for selecting the connectors with different overall performance. We discuss this in Section~\ref{subsec-framework-select-connector} and analyze the experimental results in Section~\ref{subsection-ana-vary-connector}. 
After selecting $q$ connectors, 
construction 
of the submatrix 
is shown in Algorithm~\ref{alg-sloma-heu-sub-con}.
The shortest path in Line~3 can be obtained by algorithms such as Dijkstra's algorithm.

\begin{algorithm}[ht]
\caption{Heuristic Submatrix Construction.}
\begin{algorithmic}[1]
	\REQUIRE{Observed index set $\Omega$, social graph $\mathcal{G} = (\mathcal{U}, \mathcal{E})$, number of submatrices $q$, distance threshold $d$;}
	\STATE{Obtain $q$ users as connectors from $\mathcal{U}$;} \label{alg-hsc-connectors-select}
	\FOR{$t=1,2,\dots,q$}
	\STATE{Obtain $D(u_{i_t},u_k)$: the shortest distance between $u_{i_t}$ and all the other users $u_k$s in $\mathcal{G}$;}

	\STATE{Obtain a subset of users $U^t = \big\{u_k:D(u_{i_t},u_k) \le d\big\}$;}\label{alg-hsc-users-select}
	\STATE{Obtain a subset of items $V^t = \big\{v_j:(i,j) \in \Omega, \forall u_i \in U^t \big\}$;} \label{alg-hsc-items-select}// \textit{ All items rated by users in $U^t$.}
	\STATE{Obtain $\bM^t$}; //\textit{ The $t^{th}$ submatrix based on $U^t$ and $V^t$} \label{alg-hsc-submatrix-obtain}
	\ENDFOR
	\RETURN{$\big\{\bM^t\big\}_{t=1}^q$}
\end{algorithmic}
\label{alg-sloma-heu-sub-con}
\end{algorithm}

\noindent
\textbf{Systematic Approaches.} 
The systematic approaches are based on methods for overlapping community detection, which have been intensively investigated in recent years.
We can see that social groups are equivalent to communities in social networks. 
In the literature, a community is a group of people who have more interactions within the group than those outside it~\cite{girvan2002community},
and hence users in the same community have more characteristics in common than with users outside of it. 
This leads to the low-rank property of the submatrix constructed from one community.
Naturally, communities can overlap with each other, and many works have been done for detecting such communities, 
e.g., line graph partitioning~\cite{ahn2010link}, clique percolation~\cite{palla2005uncovering}, eigenvector methods~\cite{zhang2007identification}, egonet analysis~\cite{rees2010overlapping,coscia2012demon}
and low-rank models~\cite{yang2013overlapping}.
In this work, we adopt \textit{BIGCLAM} \cite{yang2013overlapping}, 
which can scale to large datasets and has good empirical performance.
The process is shown in Algorithm~\ref{alg-sloma-sys-sub-con}.
We can see that the only difference from Algorithm~\ref{alg-sloma-heu-sub-con} is how we construct the social groups (Line~\ref{alg-hsc-connectors-select}).

\begin{algorithm}[ht]
	\caption{Systematic Submatrix Construction}
	\begin{algorithmic}[1]
		\REQUIRE{Observed index set $\Omega$, social graph $\mathcal{G} = (\mathcal{U}, \mathcal{E})$, number of submatrices $q$;}
		\STATE{Apply BIGCLAM~\cite{yang2013overlapping} to $\mathcal{G}$ to obtain the set $C$ of $q$ communities;}\label{alg-ssc-connectors-select}
		\FOR{$t=1,2,\dots,q$}
		\STATE{Obtain a subset of users $U^t \in C$;} 
		// \textit{ $U^t$ is the $t$th community.}\label{alg-ssc-users-select}
		
		\STATE{Obtain a subset of items $V^t \! = \! \big\{v_j \! : \! (i,j) \! \in \! \Omega, \forall u_i \! \in \! U^t \big\}$;} //\textit{ All items rated by users in $U^t$.} \label{alg-ssc-items-select}
		\STATE{Obtain $\bM^t$;} // \textit{ The $t^{th}$ submatrix based on $U^t$ and $V^t$} 
		
		\ENDFOR
		\RETURN{$\big\{\bM^t\big\}_{t=1}^q$}
	\end{algorithmic}
	\label{alg-sloma-sys-sub-con}
\end{algorithm}

Note that different methods lead to different numbers of users in the submatrices, 
i.e., they may cover different numbers of users in the social graph, which will influence
the overall prediction ability of SLOMA. In Sections~\ref{subsection-ana-vary-connector} and \ref{subsection-ana-systematic}, we give detailed comparisons and analysis for these different submatrix construction methods, and show that the heuristic approaches outperform the systematic ones.

\subsection{Matrix Factorization in Social Local Models}
After obtaining $q$ submatrices, we apply MF to each submatrix independently according to
Equation~(\ref{eq-mf-obj}). Note that in LLORMA, each submatrix corresponds to a local
model. Similarly, each submatrix in SLOMA corresponds to a social local model. In the
following sections, we use the terms ``local model'' and ``submatrix'' interchangeably. By
training each social local model, we can obtain $q$ groups of user latent feature matrices,
$\bU^1,\dots,\bU^q$, and item latent feature matrices, $\bV^1,\dots,\bV^q$.
Then,
the overall rating of user $u_i$ given to item $v_j$ is predicted according to the following ensemble method:
\begin{align}
\label{eq-ensemble}
\bR_{ij} = \frac{1}{q}\sum_{t=1}^{q}\bu^t_i(\bv^t_j)^{\top},
\end{align}
where $\bu^t_i$ and $\bv^t_j$, respectively, are the user-specific and item-specific latent
feature vectors in the $t$th submatrix where they occur, and $q$ is the number of
submatrices where user $u_i$ and item $v_j$ both occur. Equation~\eqref{eq-ensemble} means
that the rating of $u_i$ given to $v_j$ is predicted by the average of predicted ratings that $u_i$ gives to $v_j$ in all submatrices.

\subsection{Different Methods to Select Connectors}
\label{subsec-framework-select-connector}
The construction of social groups in SLOMA is similar to overlapping
community detection using seed set expansion~\cite{xie2013overlapping,whang2013overlapping}. There tend to be a small number of so-called long-tail users, i.e., those with few friends, which are difficult to be selected into any of the social groups, leading to incomplete coverage of SLOMA. We define two kinds of coverage in this work: i) User Coverage is the percentage of unique users in $\mathcal{U}$ selected by all of the social groups; ii) Rating Coverage is the percentage of unique ratings in $\Omega$ selected by all of the submatrices. 

Rating coverage is dependent on user coverage because when constructing each submatrix, we usually select all of the items rated by that group of users. Therefore, in this paper, we use coverage to represent any of these two concepts depending on the context.
In the local low-rank framework, the coverage of the submatrices is very important for the overall prediction performance. In LLORMA, the random selection of anchor points and choice of distance threshold are used to control the rating coverage. Due to the randomness of the anchor points, the coverage of all submatrices will be $1$ when the number of anchor points reaches a large enough value. Further, the larger the threshold is, the larger the coverage of each submatrix. 

In SLOMA, we select seeds, i.e., connectors, in the social graph and propagate from each seed along different number of hops to construct the social groups. For long-tail users who cannot be covered, we predict their ratings with the mean of all the observed ratings, which is the same approach as in~\cite{lee2016llorma}. In Section~\ref{subsection-ana-performance}, we can see that the performance of SLOMA is still better than that of LLORMA and RegSVD, demonstrating the superiority of the submatrix construction method in SLOMA. However, it is still very important to cover as many points in the rating matrix as possible. Thus, we try several connector selection methods, which are described below: 

\begin{itemize}
	\item \textbf{Hub}: Select a set of users with the largest number of neighbors.
	\item \textbf{Random}: Randomly select a set of users.
	\item \textbf{Random-Hub}: It is an integration of the above two methods, that is, we first select a larger number of hub users, e.g., 1000, and then randomly select a smaller number of the hub users, e.g., 50.
	\item \textbf{Greedy}: Each time we select a connector, we select from those not yet covered by connectors that are already selected.
\end{itemize}

After obtaining the set of connectors, we construct one submatrix around each connector user as shown in Lines~\ref{alg-hsc-users-select} and \ref{alg-hsc-items-select} of Algorithm~\ref{alg-sloma-heu-sub-con}. In Section~\ref{subsection-ana-vary-connector}, we compare the performance of these methods as well as the community-based submatrix construction method. We can see that despite the existence of a small number of uncovered users, Hub and Greedy obtain the best performance comparing to the other methods.

\vspace{-0.1in}
\subsection{SLOMA++}
In addition to constructing the submatices in SLOMA, social connections can be used to enhance each local model by applying social regularization to the factorization of each submatrix. Social regularization adds constraints that users with direct social connections should be closer in the latent space, which is introduced in Equation~\eqref{eq-soc-reg}.

For each local model, the social regularization term is defined as:
\begin{align}
\frac{\beta}{2}\sum_{i=1}^{m}\sum_{j \in \mathcal{F}(i)}\bS^t_{ij}||\bu^t_i - \bu^t_j||_2^2 ,
\label{eq-lap}
\end{align}
where $\bS^t \in \mathbb{R}^{|U^t| \times |U^t|}$ is the similarity matrix with $\bS^t_{ij}$ representing the similarity between users $u_i$ and $u_j$ in the $t$th submatrix, and $|U^t|$ is the number of users in the $t$th submatrix. Note that we apply Equation~\eqref{eq-lap} when each submatrix is factorized. We term this model SLOMA++.

To calculate the similarities between two users $u_i$ and $u_j$, we utilize the popular Person Correlation Coefficient (\textit{PCC})~\cite{breese1998empirical}:
\begin{equation*}
\bS^t_{ij} \!=\! \frac{\sum\limits_{f \in T(i,j)} (\bM^t_{if} - \bar{M}^t_i) (\bM^t_{jf} - \bar{M}^t_j)}
{\sqrt{\sum\limits_{f \in T(i,j)}(\bM^t_{if} - \bar{M}^t_i)^2} \sqrt{\sum\limits_{f \in T(i,j)}(\bM^t_{jf} - \bar{M}^t_j)^2}},\!\!\!\!
\end{equation*}
where $\bar{M}^t_i$ represents the mean value of all of the ratings of user $u_i$ in the $t$th submatrix $\bM^t$, and $I^t(i)$ represents the set of items rated by user $u_i$ in the $t$th submatrix $\bM^t$. $T(i,j) = I^t(i) \cap I^t(j)$, representing the set of common items rated by $u_i$ and $u_j$ in the $t$th submatrix. Note that when constructing the submatrix for a group of users, we take all of the group's ratings in $\bO$. Thus, for the same user pair $u_i$ and $u_j$, $T(i,j)$ will be the same across all submatrices. This is why we remove the superscript $t$ from $T(i,j)$. Further, we employ a mapping function $g(x) = \frac{1}{2} (x+1)$ to bound the range of \textit{PCC} similarity to $[0, 1]$.

For each submatrix, we need to solve the following optimization problem:
\begin{align}
\min_{\bU,\bV}\;&\;\frac{1}{2}\sum_{(i,j) \in \Omega}(\bO_{ij} - \bu_i\bv_j^{\top})^2\nonumber + \frac{\lambda}{2}\left( ||\bU||_F^2 + ||\bV||_F^2 \right)\nonumber\\
&+ \frac{\beta}{2}\sum_{i=1}^{m}\sum_{j \in \mathcal{F}(i)}\bS_{ij}||\bu_i - \bu_j||_2^2
\label{eq-sloma++-obj}.
\end{align}
Note that we leave out the superscript $t$ for simplicity of notations. From this equation we can see that when $\beta = 0$, we recover SLOMA. Equation~\eqref{eq-sloma++-obj} can be solved by gradient descend methods \cite{mnih2007probabilistic,koren2008factorization}.

\section{Experiment}
\label{sec-exp}
In this section, we introduce the details of the experiments, including the datasets, the evaluation metrics, and the baselines.

\subsection{Datasets}
We conduct the experiments on two real-world datasets: Yelp and Douban. Yelp is a location-based website where users can give ratings to and write reviews on items like restaurants, theaters, and businesses. The Yelp dataset is provided by the Yelp Dataset Challenge,\footnote{https://www.yelp.com/dataset\_challenge} which has been used in previous research in recommendation \cite{lee2016llorma,li2015overlapping,mcauley2013hidden}.
Douban is a Chinese website where users can rate and share their opinions on items such as movies and books. The Douban Dataset is obtained from~\cite{ma2011recommender}. 
In both websites, users can build Facebook-style connections to each other and all ratings are in the range 1 to 5. Therefore, the datasets are ideal for evaluating the effectiveness of our proposel model.

We preprocess the two datasets by removing users without any friends or with ratings fewer than 5. These users are called, respectively, ``social cold-start'' and ``cold start'' users in RSs.
The statistics of the two preprocessed datasets are shown in Table~\ref{tb-dataset-stat}. Note that in the table, R\_density represents the density of the rating matrix, and S\_edges and S\_density represent the social connections and density of the social matrix, respectively.

\begin{table*}[ht]
	\centering
	\caption{Statistics of Datasets ($R\_density=\frac{\#Ratings}{\#Users\times \#Items}$, $S\_density=\frac{\#2 \times S\_edges}{\#Users\times \#Users}$).}
	\vspace{-4px}
	\label{tb-dataset-stat}
	\begin{tabular}[\columnwidth]{c|cccc|cc}
		\hline
		& Users & Items & Ratings & R\_density & S\_edges & S\_density \\ \hline
		Yelp &  76,220  & 79,257  & 1,352,762   &0.022\% & 647,451 &0.022\% \\ \hline
		Douban & 103,054  & 57,908  & 15,129,113 &  0.254\% & 753,358 &0.028\% \\ \hline
	\end{tabular}
\end{table*}

\subsection{Evaluation Metrics}
We choose two evaluation metrics, Mean Absolute Error (MAE) and Root Mean Square Error (RMSE).
They are defined as
\begin{align*}
\text{MAE} 
& = \frac{1}{|\bar{\Omega}|} \sum_{(i,j) \in \bar{\Omega}}{|\bO_{ij} - \bR_{ij}|},
\\
\text{RMSE} 
& = \sqrt{\frac{1}{|\bar{\Omega}|} \sum_{(i,j) \in \bar{\Omega}}{(\bO_{ij} - \bR_{ij})^2}},
\end{align*} 
where $\bar{\Omega}$ is the set of all user-item pairs $(i,j)$ in the test set,
and $\bR_{ij}$ is the corresponding rating predicted by the algorithm.
These metrics are popular for the task of rating prediction in the  literature~\cite{mnih2007probabilistic,koren2008factorization,ma2011recommender}.

\subsection{Experimental Settings}
We compare our proposed models with the following state-of-the-art methods:
\begin{itemize}
	\item \textbf{RegSVD}~\cite{paterek2007improving}: It is the standard matrix
	factorization method with $\ell_2$ regularization. We implement it according to~\cite{paterek2007improving}.
	\item \textbf{LLORMA}~\cite{lee2016llorma}: It is based on the local low-rank assumption, which SLOMA also assumes. We implement it according to \cite{lee2016llorma}.
	\item \textbf{SocReg}~\cite{ma2011recommender}: It is a state-of-the-art method that integrates social connections into MF by employing social connections as regularization terms. We implement it according to \cite{ma2011recommender}.
	\item \textbf{SLOMA}: This is our proposed model. It utilizes social connections among users to form social groups for local low-rank factorization.
	\item \textbf{SLOMA++}: This is the same as SLOMA except that it applies social regularization to the factorization
	of submatrices, as is shown in Equation~\eqref{eq-lap}.
\end{itemize}

Following the experimental settings in~\cite{lee2016llorma},
we also randomly split each dataset into training and test data with a ratio of 8:2. 
In the training process, the training data are used to fit the model, and the test data are
used to calculate the prediction errors of the models. We repeat each experiment five times
by randomly splitting the datasets and report the average results. 

\begin{table*}[]
	\centering
	\caption{Performance of different methods in Yelp and Douban with $K=10,20$. The best two results are highlighted. Improvements are measured by the reduction of RMSE comparing to SLOMA++.}
	\vspace{-5px}
	\label{tb-performance}
	\begin{tabular}{c|c|c|C{45px}C{45px}C{45px}C{45px}C{45px}}
		\hline
		\multicolumn{1}{l|}{Datasets} & K & Metrics & RegSVD & LLORMA & SocReg & SLOMA & \multicolumn{1}{l}{SLOMA++} \\ \hline
		\multirow{8}{*}{Yelp} & \multirow{4}{*}{10} & MAE & 0.9478 & 0.9459 & {\ul \textbf{0.9228}} & 0.9362 & {\ul 0.9301} \\ 
		&  & Improve &  +1.87\% & +1.67\% & -0.79\% & +0.65\%  &  \\
		&  & RMSE & 1.1908 & 1.1843 & 1.1802 & {\ul 1.1760} & {\ul \textbf{1.1755}} \\ 
		&  & Improve &  +1.28\% & +0.74\% & +0.40\% & +0.04\%  &  \\\cline{2-8} 
		& \multirow{4}{*}{20} & MAE & 0.9499 & 0.9477 & {\ul 0.9190} & 0.9389 & {\ul 0.9240} \\		
		&  & Improve &  +2.73\% & +2.50\% & -0.54\% & +1.59\%  &  \\
		&  & RMSE & 1.1918 & 1.1862 & {\ul 1.1754} & 1.1788 & {\ul \textbf{1.1698}} \\
		&  & Improve &  +1.85\% & +1.38\% & +0.48\% & +0.76\%  &  \\\hline
		\multirow{8}{*}{Douban} & \multirow{4}{*}{10} & MAE & 0.5828 & 0.5811 & {\ul 0.5662} & 0.5744 & {\ul \textbf{0.5603}} \\ 
		&  & Improve &  +3.86\% & +3.58\% & +1.04\% & +2.45\%  &  \\
		&  & RMSE & 0.7347 & 0.7310 & {\ul 0.7165} & 0.7255 & {\ul \textbf{0.7105}} \\ 
		&  & Improve &  +3.29\% & +2.80\% & +0.84\% & +2.07\%  &  \\\cline{2-8}		
		& \multirow{4}{*}{20} & MAE & 0.5803 & 0.5779 & {\ul 0.5638} & 0.5715 & {\ul \textbf{0.5573}} \\
		&  & Improve &  +3.96\% & +3.56\% & +1.15\% & +2.48\%  &  \\
		&  & RMSE & 0.7320 & 0.7278 & {\ul 0.7142} & 0.7225 & {\ul \textbf{0.7080}} \\
		&  & Improve & +3.28\% & +2.72\% & +0.87\% & +2.01\%  &  \\\cline{3-8}		
		\hline
	\end{tabular}
\end{table*}

\section{Analysis}
\label{sec-analysis}
In this section, we present and compare the experimental results of our proposed SLOMA and SLOMA++ against the baselines. We try to answer the following questions:
\begin{enumerate}
\item[(1).] What is the recommendation performance of our proposed models
comparing to the state-of-the-art methods?\label{ana-question-1}

\item[(2).] How do the parameters in our models affect the overall performance?
Specifically, how do the number of submatrices and hops, i.e., distance threshold, affect recommendation performance?

\item[(3).] What is the performance of different methods of connector selection?

\item[(4).] Which submatrix construction approach, heuristic or systematic, is better for SLOMA and SLOMA++?
\end{enumerate}

\subsection{Recommending Performance}
\label{subsection-ana-performance}

To answer Question~(1), we show the recommendation performance of the baselines and our
proposed methods in Table~\ref{tb-performance}. We can see that SLOMA++ consistently
outperforms all the other methods under different $K$s in both datasets. This demonstrates the efficacy of integrating social connections with local low-rank framework. When comparing RegSVD, LLORMA, and SLOMA, we observe that SLOMA has better performance than LLORMA and RegSVD, which can be attributed to the better submatrices SLOMA builds with the help of social connections. When further studying the results of SocReg, SLOMA and SLOMA++, we can see that SLOMA++ is the best. This demonstrates that SLOMA++ makes the best use of social connections. Taking advantage from both the local low-rank assumption and social recommendation, SLOMA++ beats SocReg with the local low-rank framework and SLOMA with social regularization. Through these two comparisons, we can conclude that both social regularization and local low-rank can help improve the recommending performance.

When comparing SLOMA and SocReg, we can see that the performance of SocReg is consistently a bit better than that of SLOMA. This result not only indicates the power of social regularization but also points to a limitation of SLOMA, that is, it does not cover the long-tail users very well when constructing the submatrices and hence can only predict the long-tail users' ratings based on the mean of all the existing ratings. The latter also explains the phenomenon that the performance gain of SLOMA against RegSVD is limited. Therefore, it is very important to design better submatrix construction methods with larger coverage. We leave this for future research.

When comparing the performance on Yelp and Douban, we can see that the improvement on Douban is more significant than that on Yelp, which may be attributed to Douban's larger density of social connections (see Table~\ref{tb-dataset-stat}). It means that better performance can be obtained when social networks are dense. This is intuitive because the more friends a user has, the more likely his or her behaviors are affected by friends.

\vspace{-0.1in}
\subsection{Impact of the Number of Local Models}
\begin{figure}[!ht]
	\centering
	\subfigure[$K=10$@Yelp.]{\includegraphics[width=0.24\textwidth]{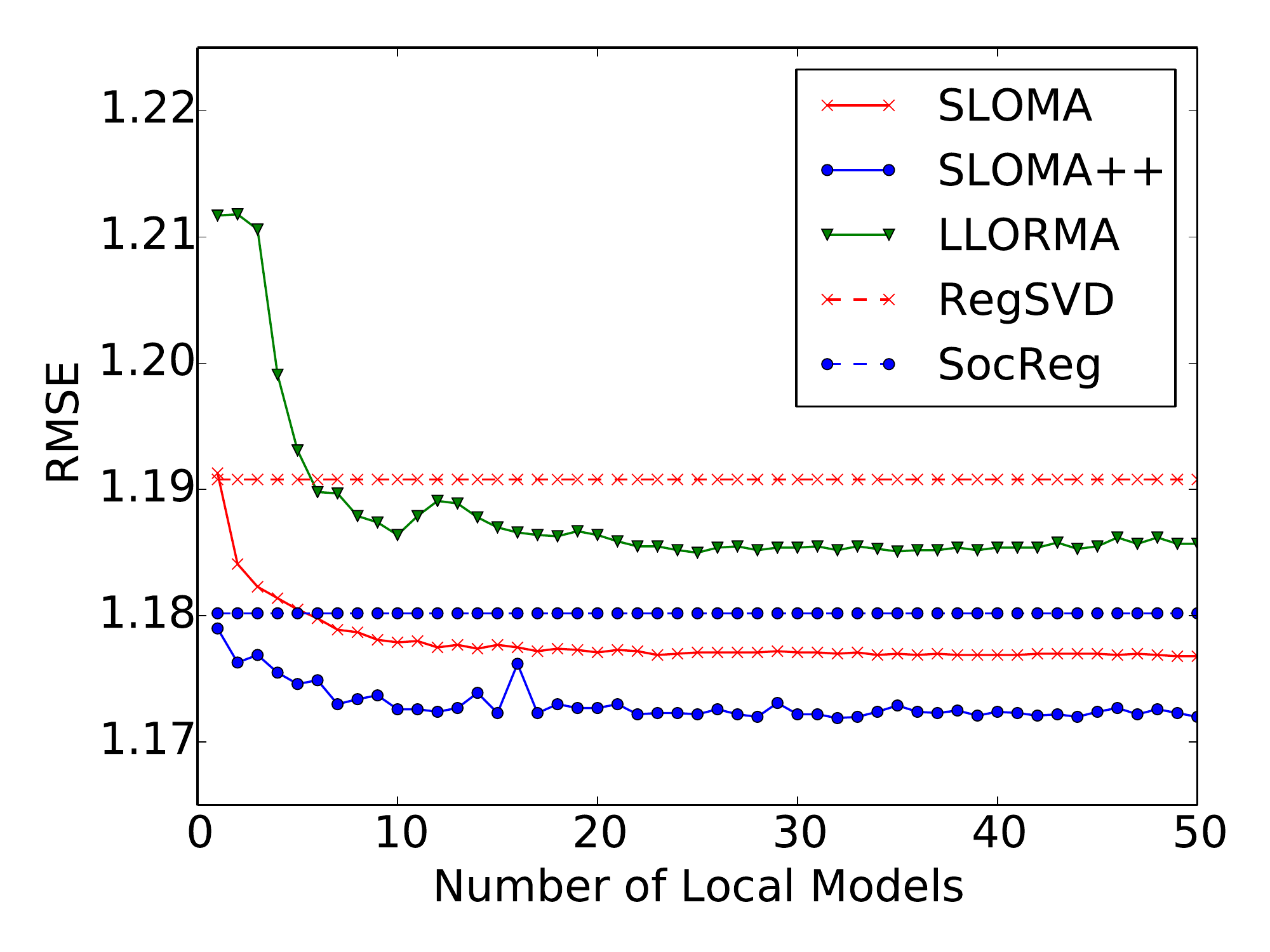}}
	\subfigure[$K=20$@Yelp.]{\includegraphics[width=0.24\textwidth]{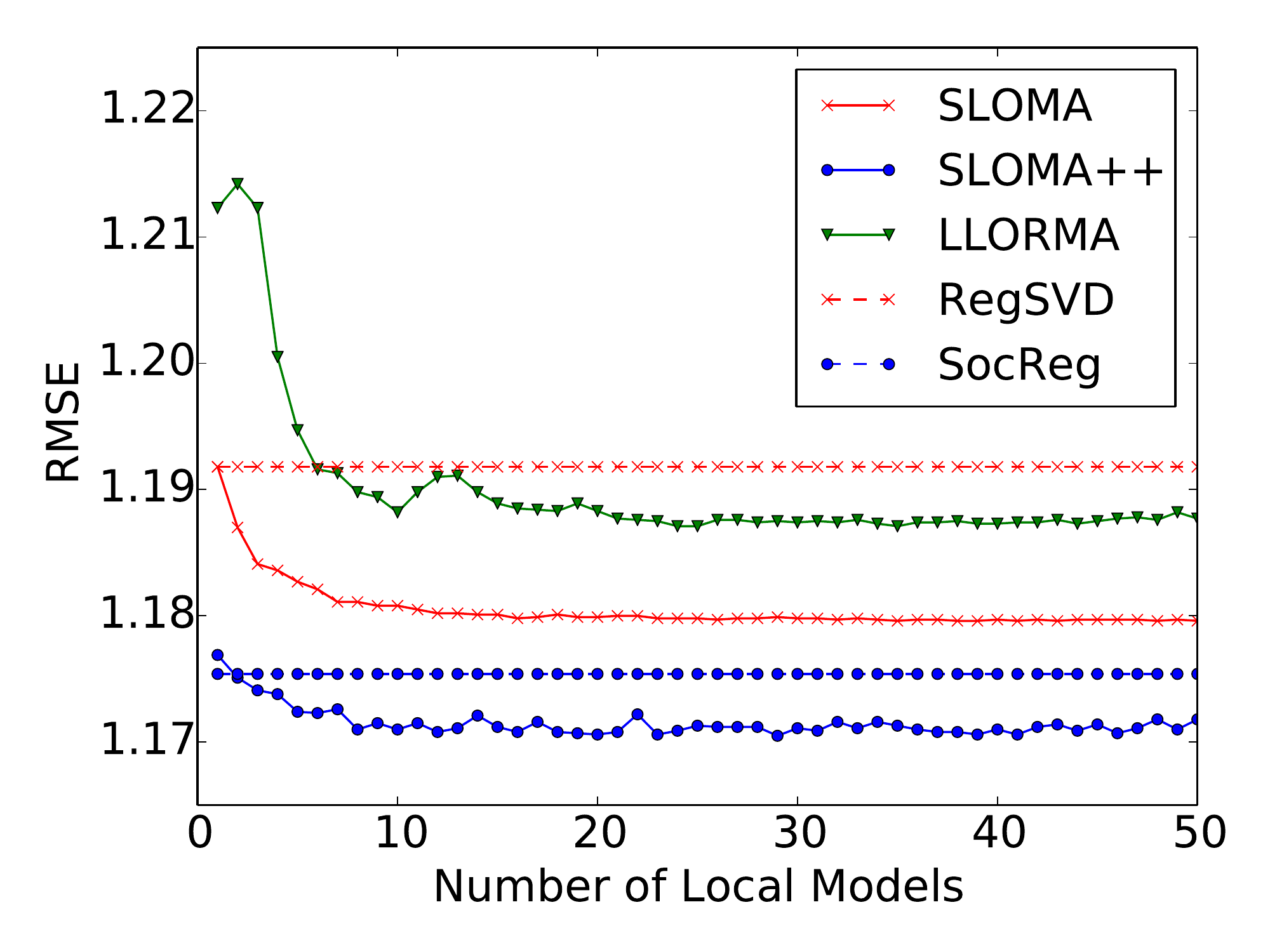}}
	\subfigure[$K=10$@Douban.]{\includegraphics[width=0.24\textwidth]{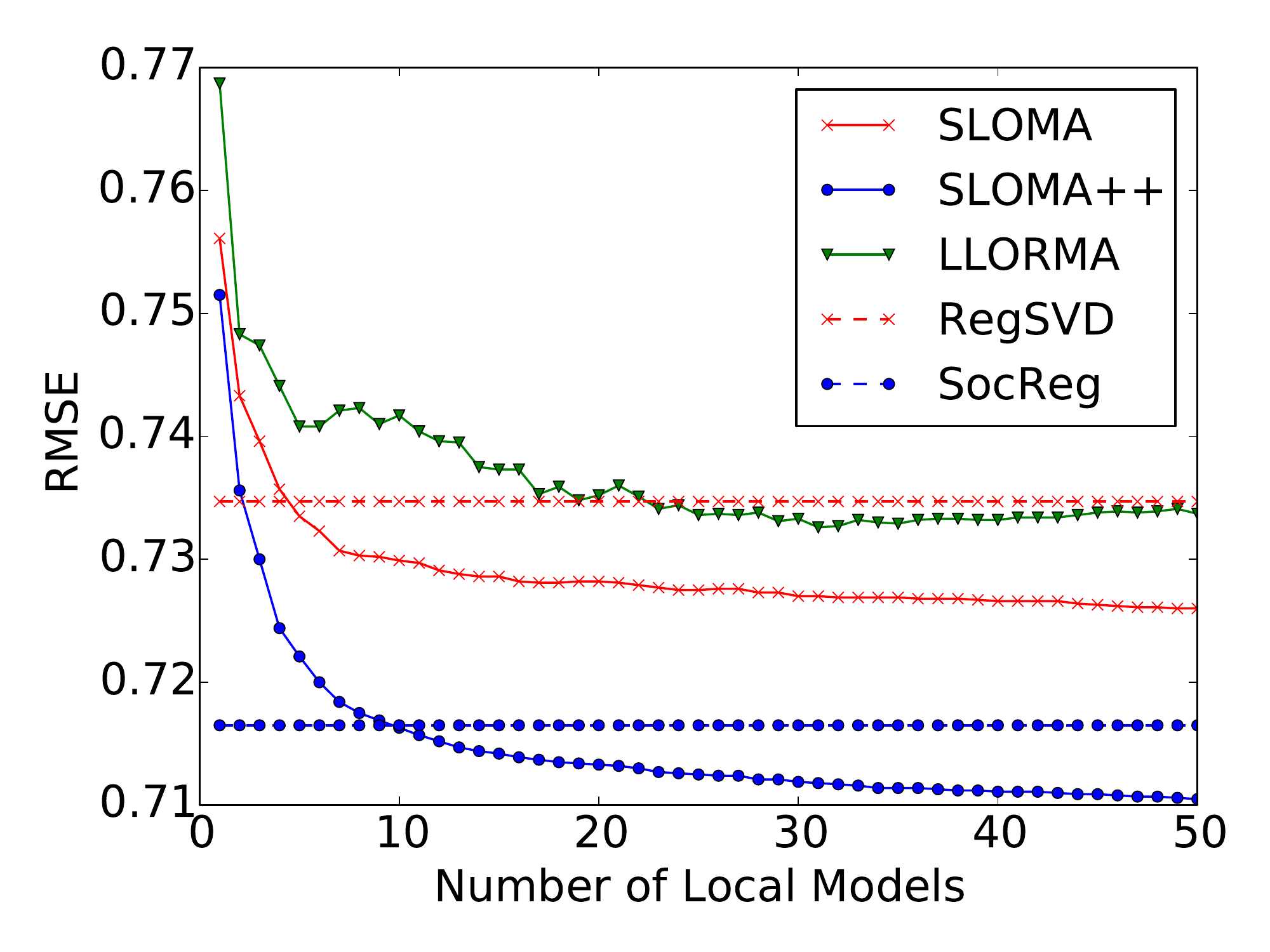}}
	\subfigure[$K=20$@Douban.]{\includegraphics[width=0.24\textwidth]{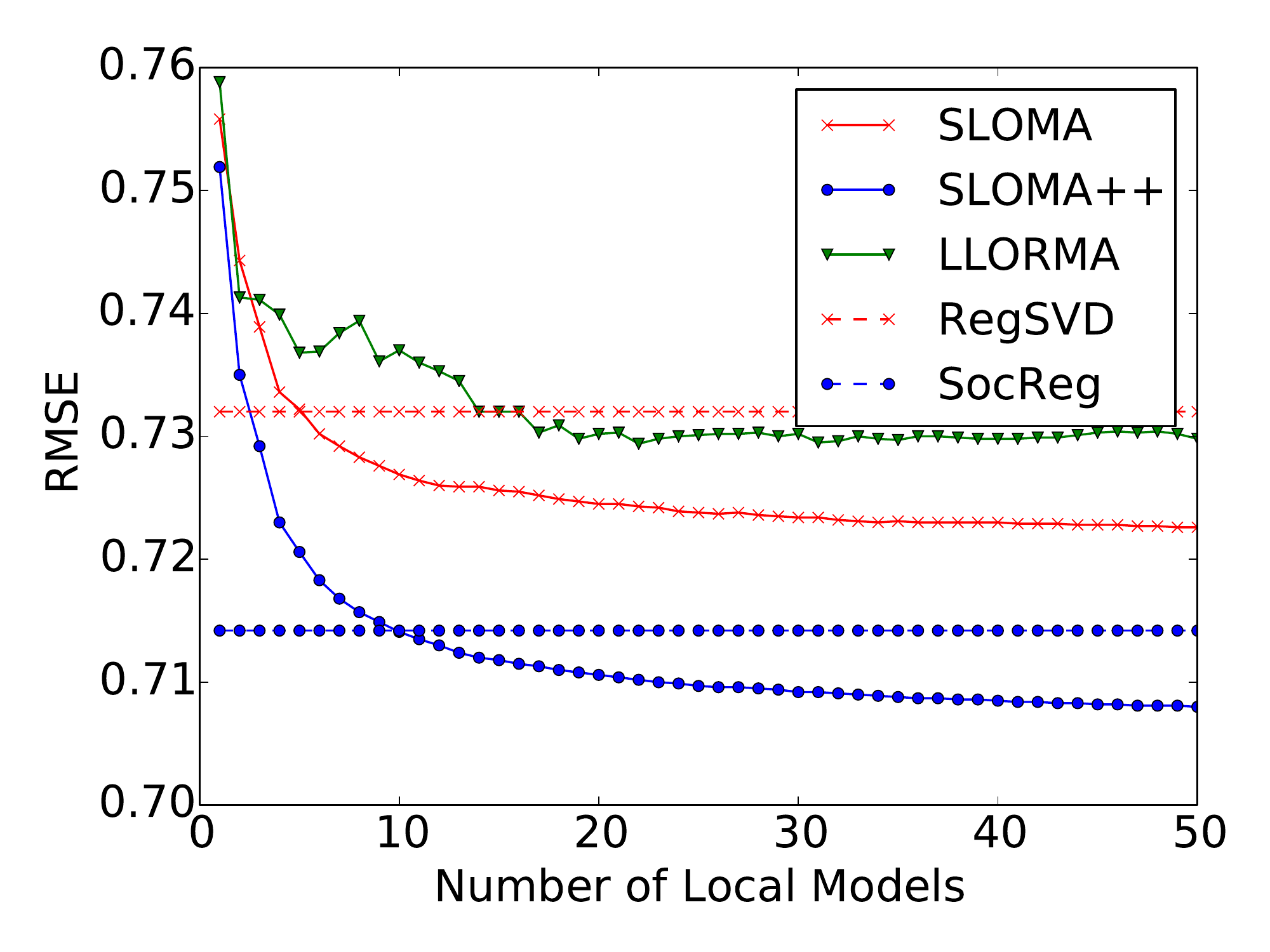}}
	\centering
	\vspace{-5px}
	\caption{RMSEs vs different numbers of local models on Yelp and Douban.}
	\label{fig-vary-local-number}
\end{figure}

In this section, we show how performance varies with different numbers of local models in
LLORMA, SLOMA, SLOMA++. The results are shown in Figure~\ref{fig-vary-local-number}. We also
plot the RMSEs of RegSVD and SocReg for comparison. From the figures, we can see that with
increasing number of local models, RMSE's of LLORMA, SLOMA, and SLOMA++ all decrease. This means that the more local models, the better performance we can get. However, when the number is large enough, e.g., around 30, the performance gains become marginal. This trend is consistent with the results reported in~\cite{lee2016llorma}, which indicates that SLOMA and SLOMA++ show similar performance trends to LLORMA with the number of local models varying. 

When the number of local models is smaller than $10$, the performance is not as good as what we can achieve at $30$. The reason is that fewer submatrices means less overlaps of the local models, which impairs the ensemble prediction accuracy according to Equation~\eqref{eq-ensemble}. Further, it also reduces the overall coverage of SLOMA and SLOMA++, which means more users' ratings are predicted by the mean of other users instead of by the learned model.

Again we can see that SLOMA++ consistently outperforms all the other baselines when the local model is large enough, e.g., greater than 10. In practice, it is good enough to set the number of local models to $50$, which provides good prediction ability while avoiding higher costs of space and computation.
\vspace{-0.2in}
\vspace{-0.5in}
\begin{figure}[!ht]
	\subfigure[$K=10$@Yelp.]{\includegraphics[width=0.24\textwidth]{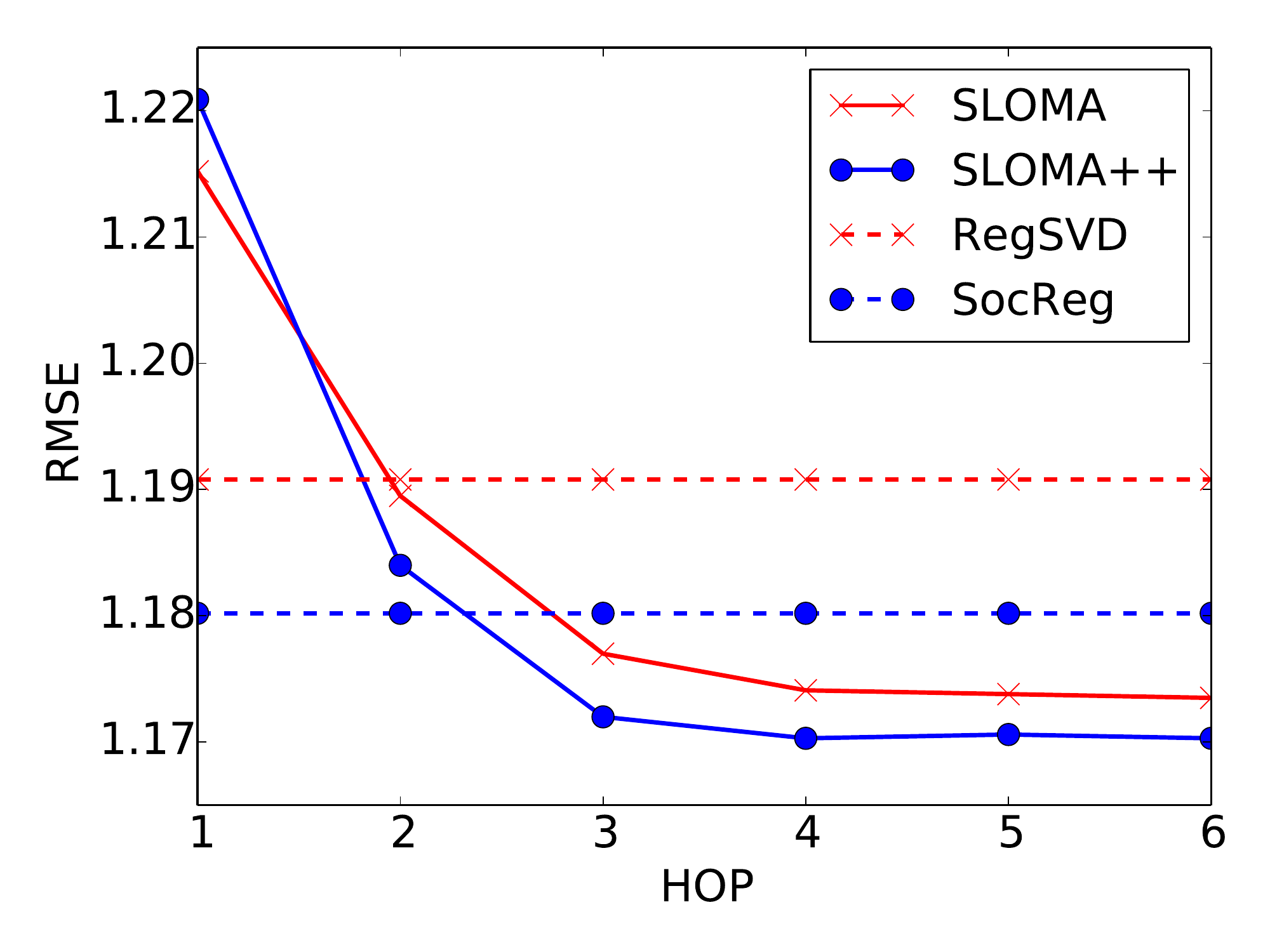}\label{fig-ipd-a}} 
	\subfigure[$K=20$@Yelp.]{\includegraphics[width=0.24\textwidth]{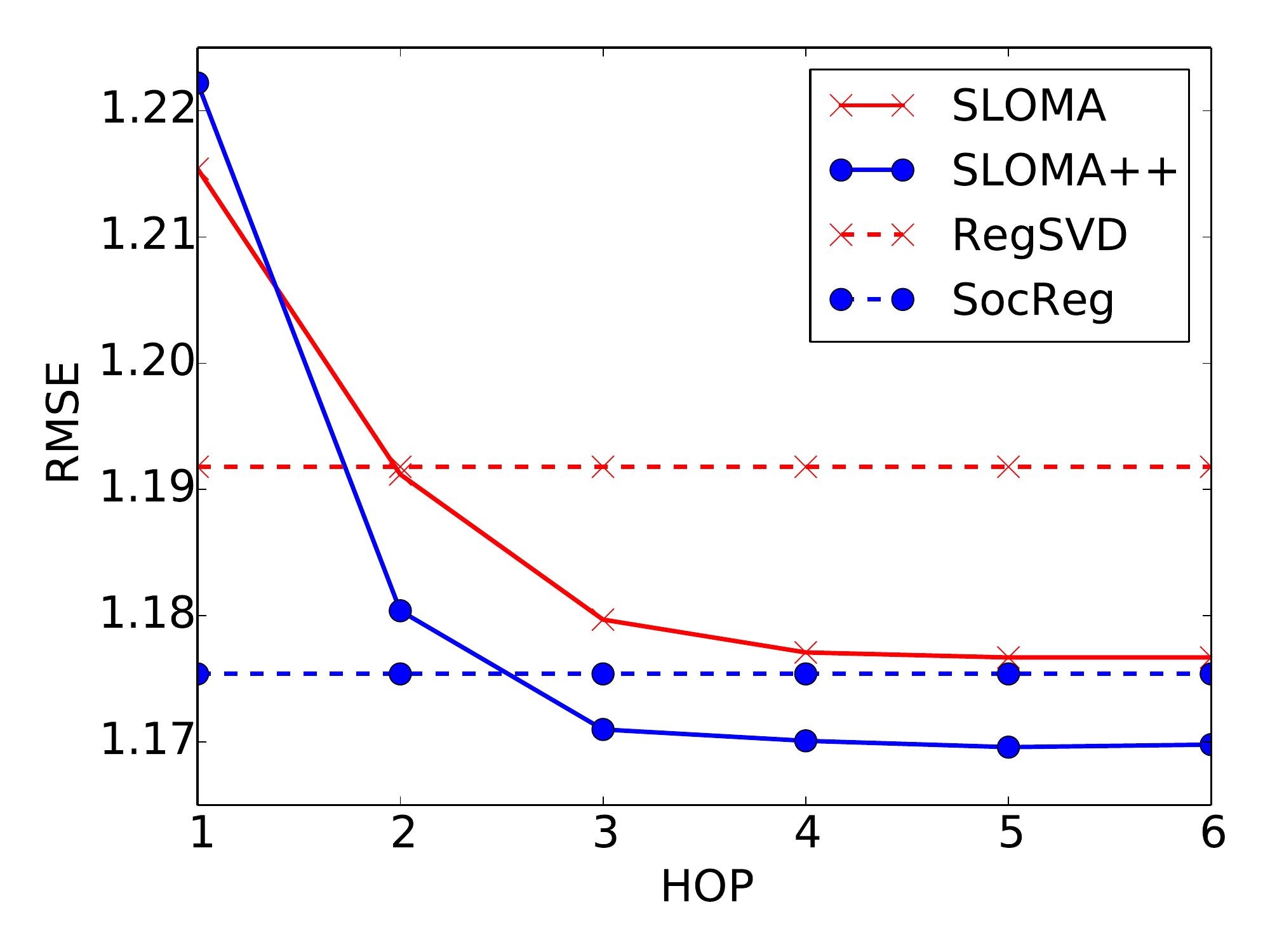}\label{fig-ipd-b}}
	\subfigure[$K=10$@Douban.]{\includegraphics[width=0.24\textwidth]{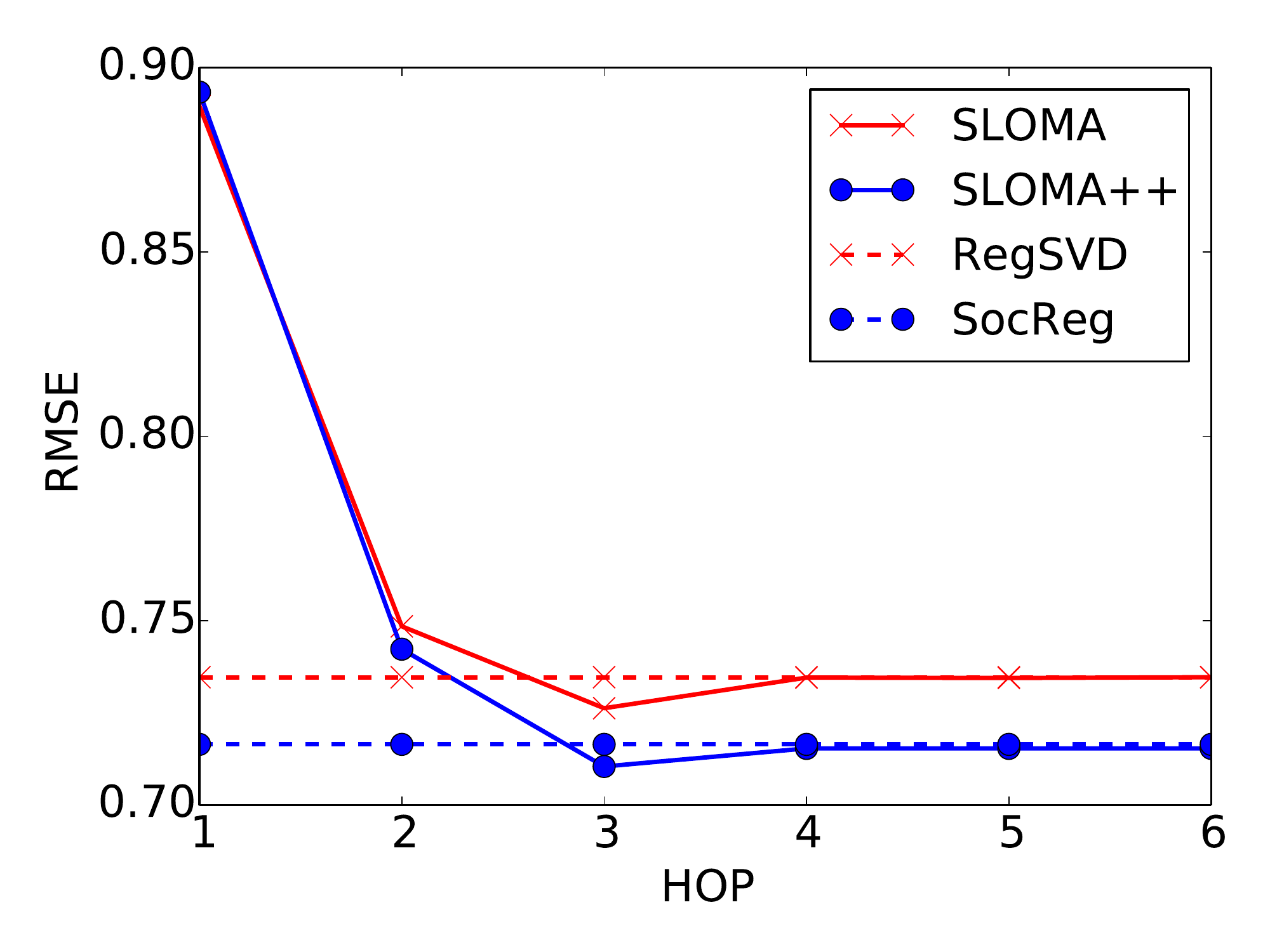}\label{fig-ipd-c}}
	\subfigure[$K=20$@Douban.]{\includegraphics[width=0.24\textwidth]{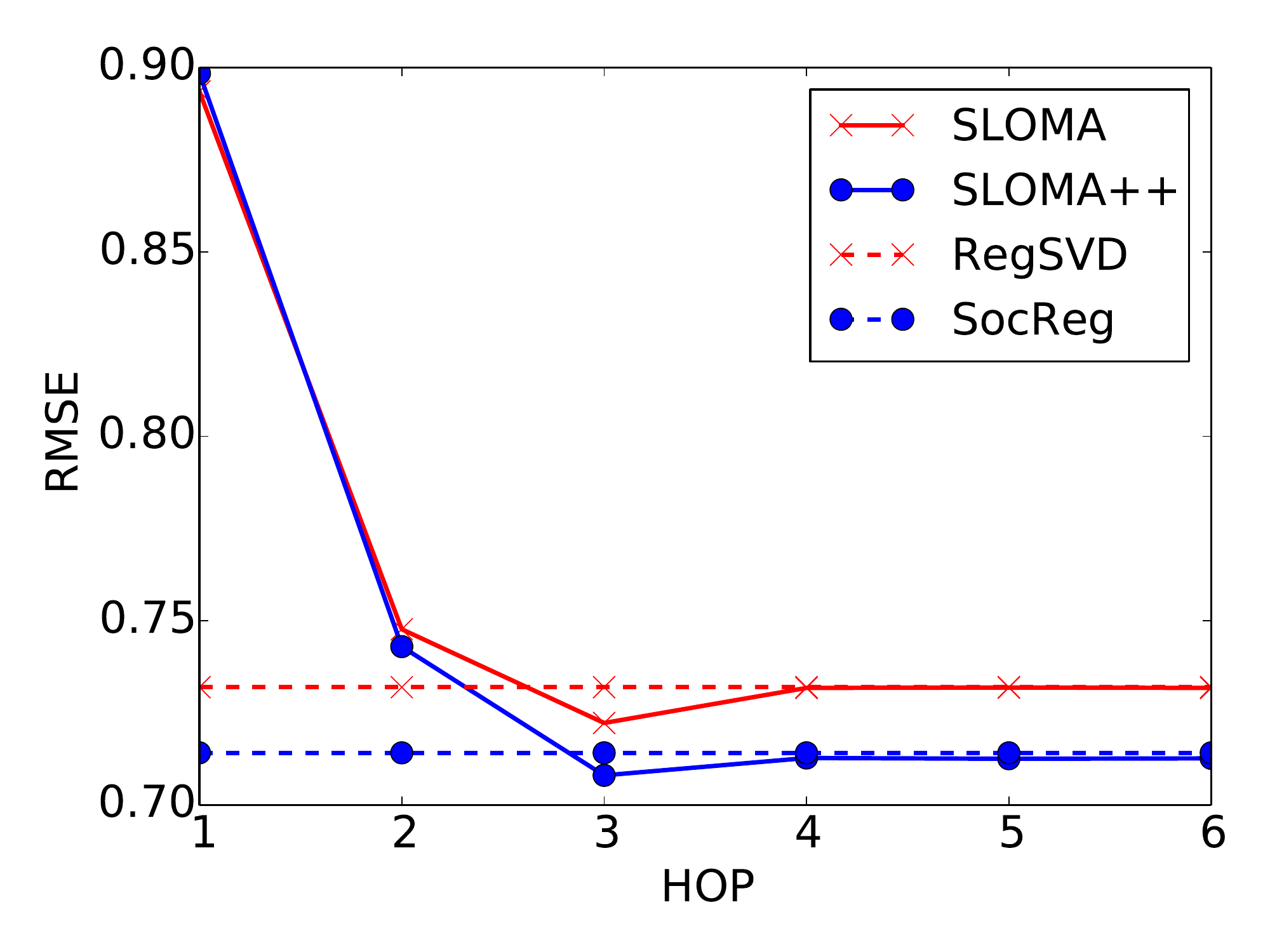}\label{fig-ipd-d}}
	\vspace{-5px}
	\caption{RMSEs vs different hops on Yelp and Douban.}
	\label{fig-vary-degree}
\end{figure}

\subsection{Impact of the Number of Hops}
\label{ana-ipd}
In SLOMA and SLOMA++, we firstly select a small number of connectors, and then take all of
their friends within different hops, i.e., the distance threshold $d$ in
Algorithm~\ref{alg-sloma-heu-sub-con}, in the social network to construct the social groups.
Therefore, different hops mean different numbers of users included in each social group,
thus different coverages of each local model, which results in different prediction
capabilities of the proposed models. Due to the small-world phenomenon in social networks~\cite{travers1967small,kleinberg2000small}, 
which states that two people can be connected by a small number of intermediates in social
networks, the so-called Six Degree Separation. Thus, we limit the hop to the range $[1,6]$
in the experiments. 

Results are shown in Figure~\ref{fig-vary-degree}.
As can be seen, the trend of the performance w.r.t. the hop is very similar to that of the
number of local models, which is, the performance becomes better with increasing of hops and
then becomes stable when the hop is large enough. Specifically, when $hop < 3$, the
performance of SLOMA and SLOMA++ is worse than those of RegSVD and SocReg. The reason is
that the coverage of each local model is not large enough, leading to a decrease in overall
coverage of all local models. Therefore, the ensemble ability is impaired. However, when
$hop \geq 3$, the performance of SLOMA and SLOMA++ becomes better, and stable when hop keeps
increasing. Therefore, it means that a larger hop is useful for SLOMA and SLOMA++. However, it does not have to be very large because it will lead to higher space and computation cost but only marginal performance gain.

An interesting observation is that on Douban, at $hop > 3$, the performance of SLOMA and
SLOMA++ decreases to a level that is very close to those of RegSVD and SocReg. This means that the benefit of the local low-rank assumption vanishes. We analyze the experimental results and find that SLOMA and SLOMA++ can cover all the users in Douban, thus each local model is reduced to the same one as RegSVD. Specifically, the variance of each local model decreases when the hop is too large, thus impairing the ensemble performance of all local models. Therefore, in the previous experiments comparing SLOMA and SLOMA++ against other baselines, we set $d=3$ when constructing submatrices according to Algorithm~\ref{alg-sloma-heu-sub-con}. 

In summary, Question~(2) can be understood clearly based on the above two sections.

\begin{figure*}[ht]
\centering
\subfigure[SLOMA($K=10$@Yelp).]{\includegraphics[width=0.24\textwidth]{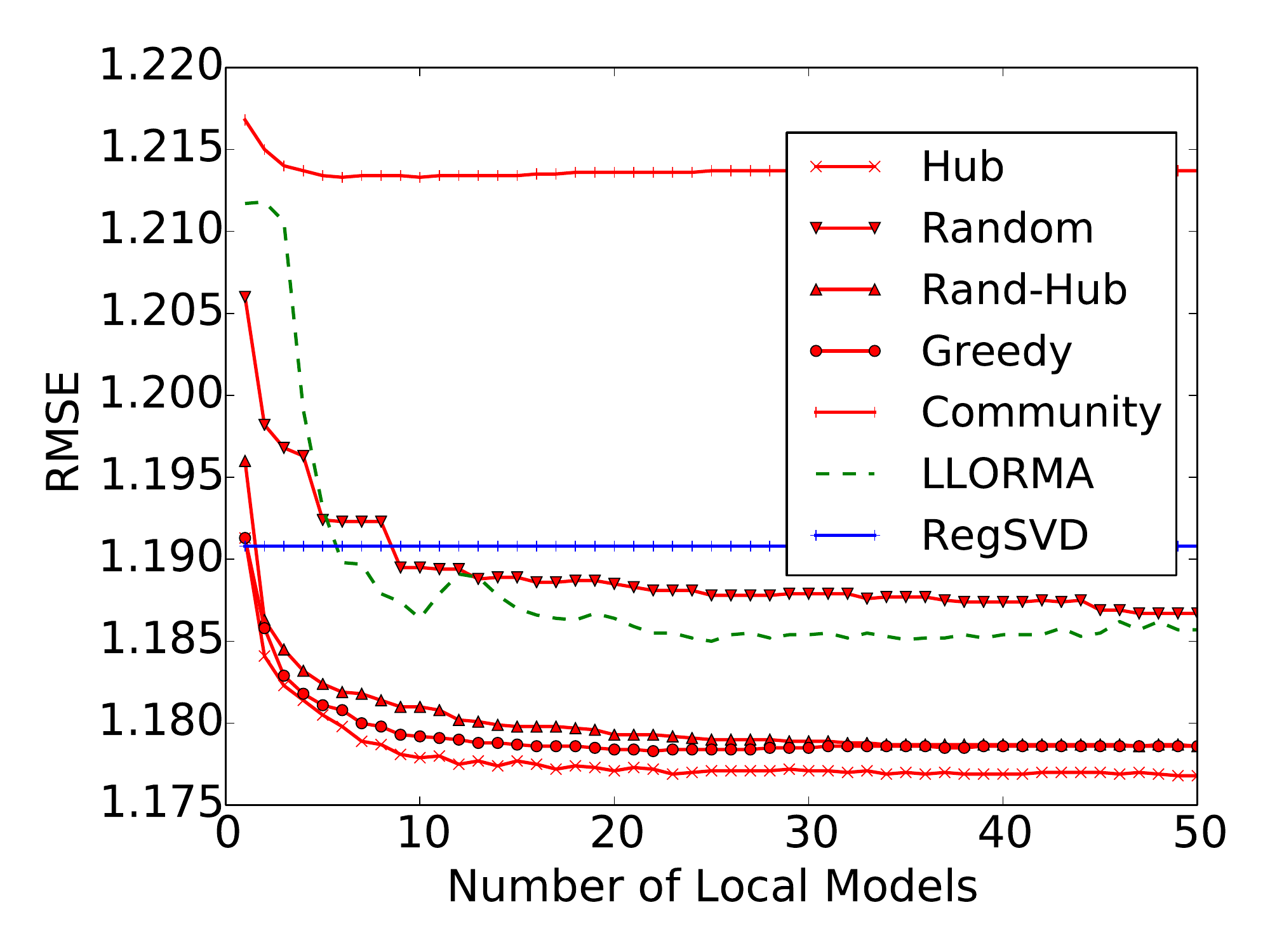}\label{fig-sloma-yelp-K10}}
\subfigure[SLOMA($K=20$@Yelp).]{\includegraphics[width=0.24\textwidth]{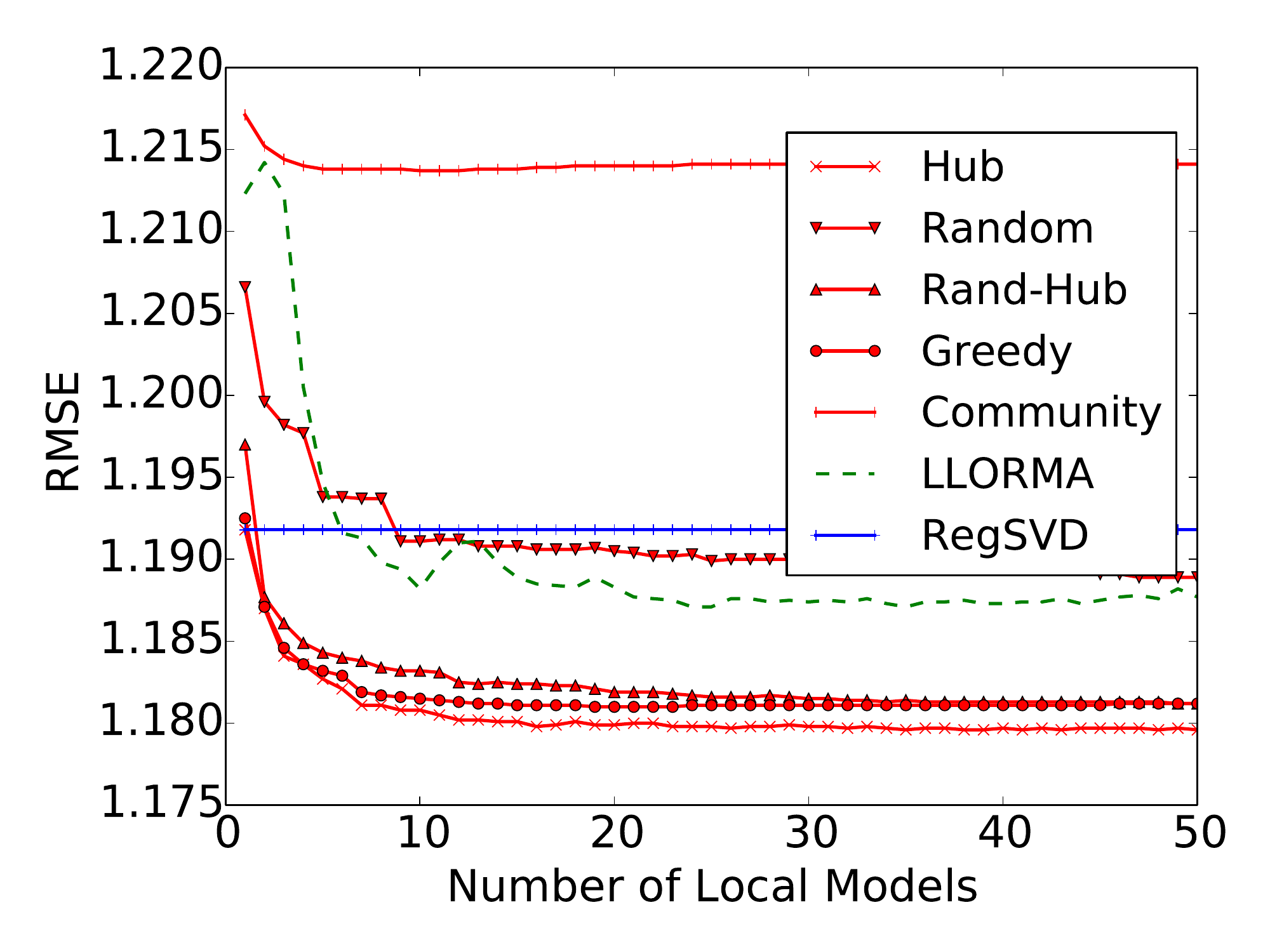}\label{fig-sloma-yelp-K20}}
\subfigure[SLOMA++($K=10$@Yelp).]{\includegraphics[width=0.24\textwidth]{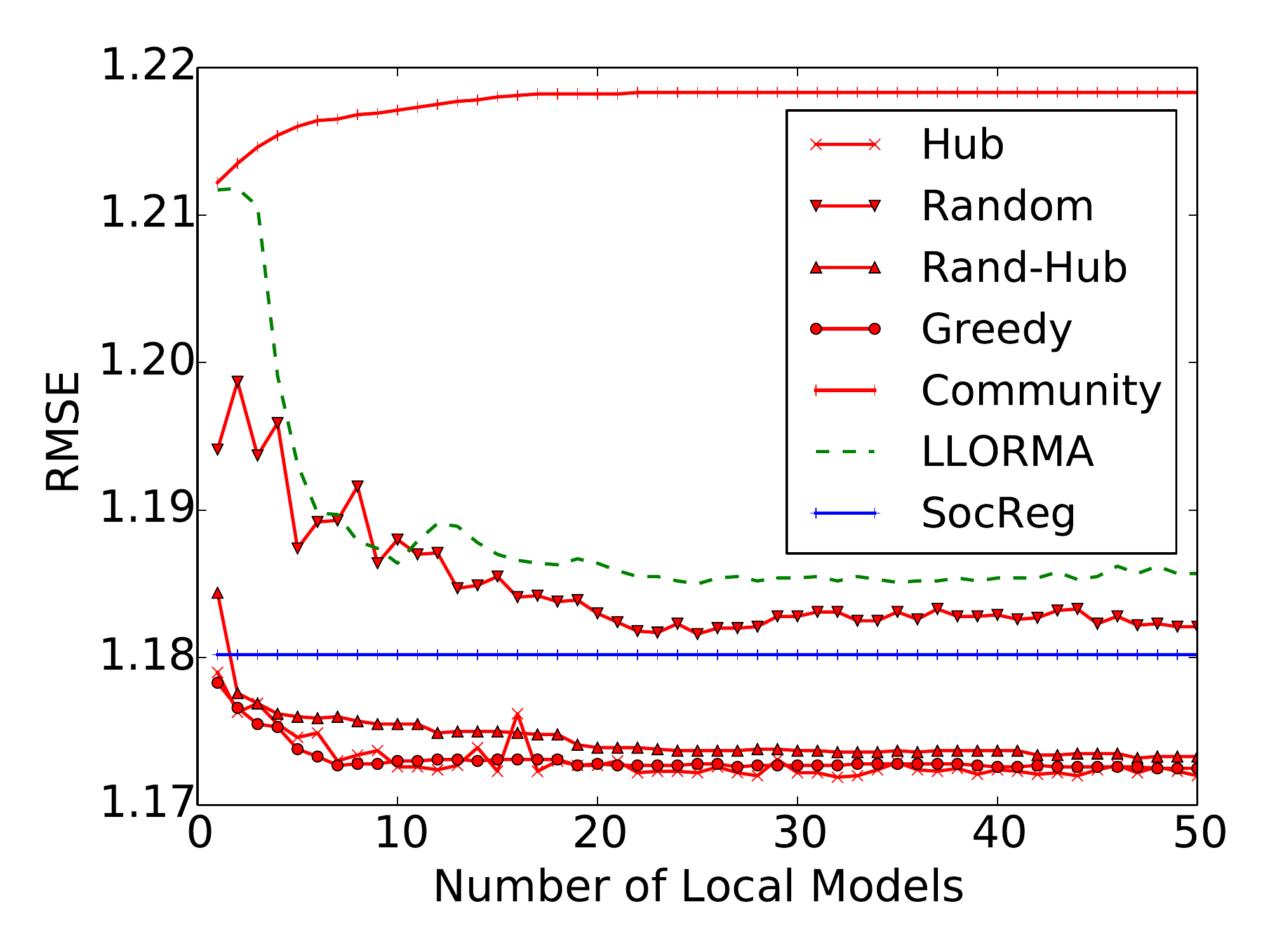}\label{fig-sloma++-yelp-K10}}
\subfigure[SLOMA++($K=20$@Yelp).]{\includegraphics[width=0.24\textwidth]{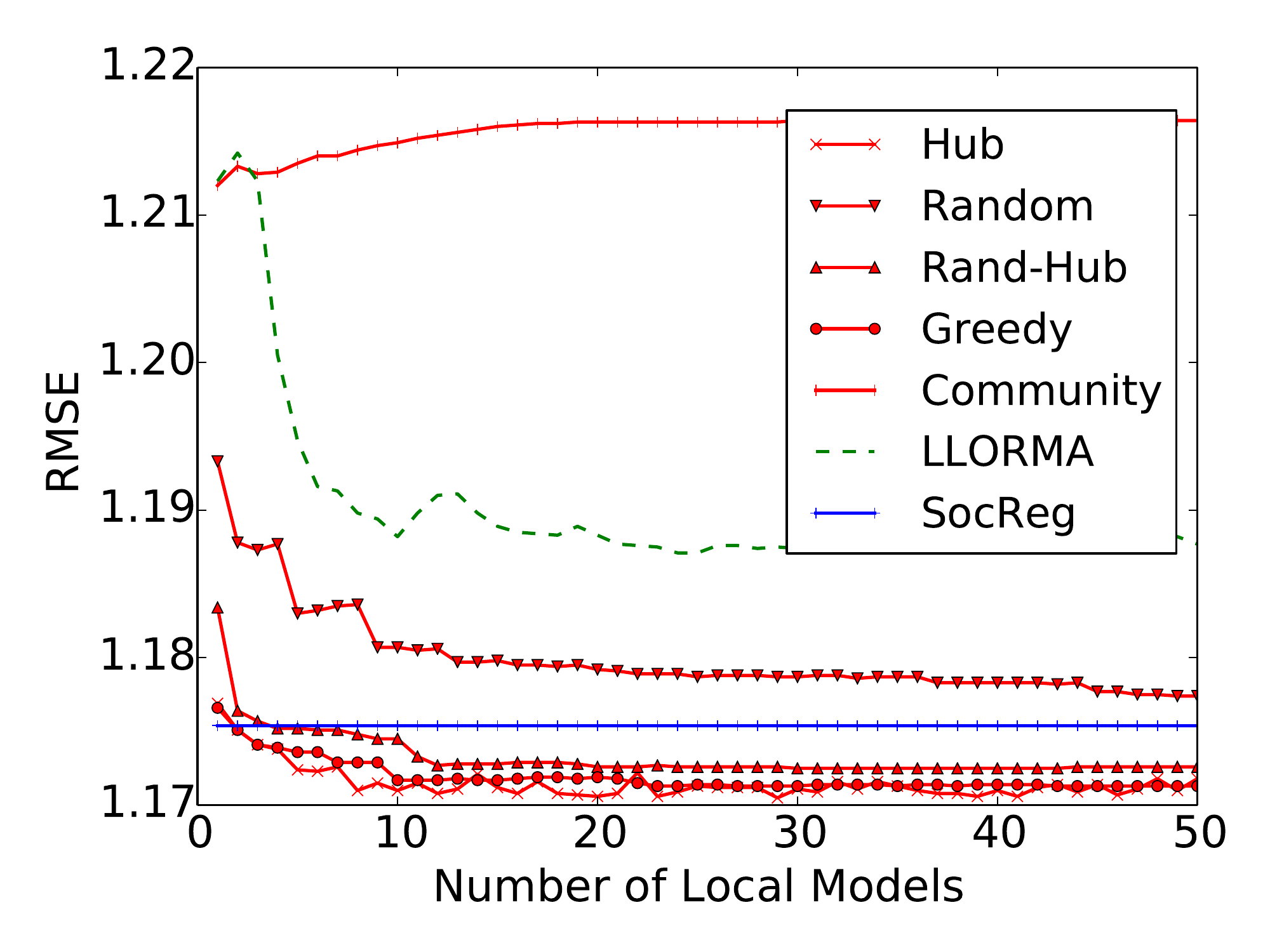}\label{fig-sloma++-yelp-K20}}\\
\subfigure[SLOMA($K=10$@Douban).]{\includegraphics[width=0.24\textwidth]{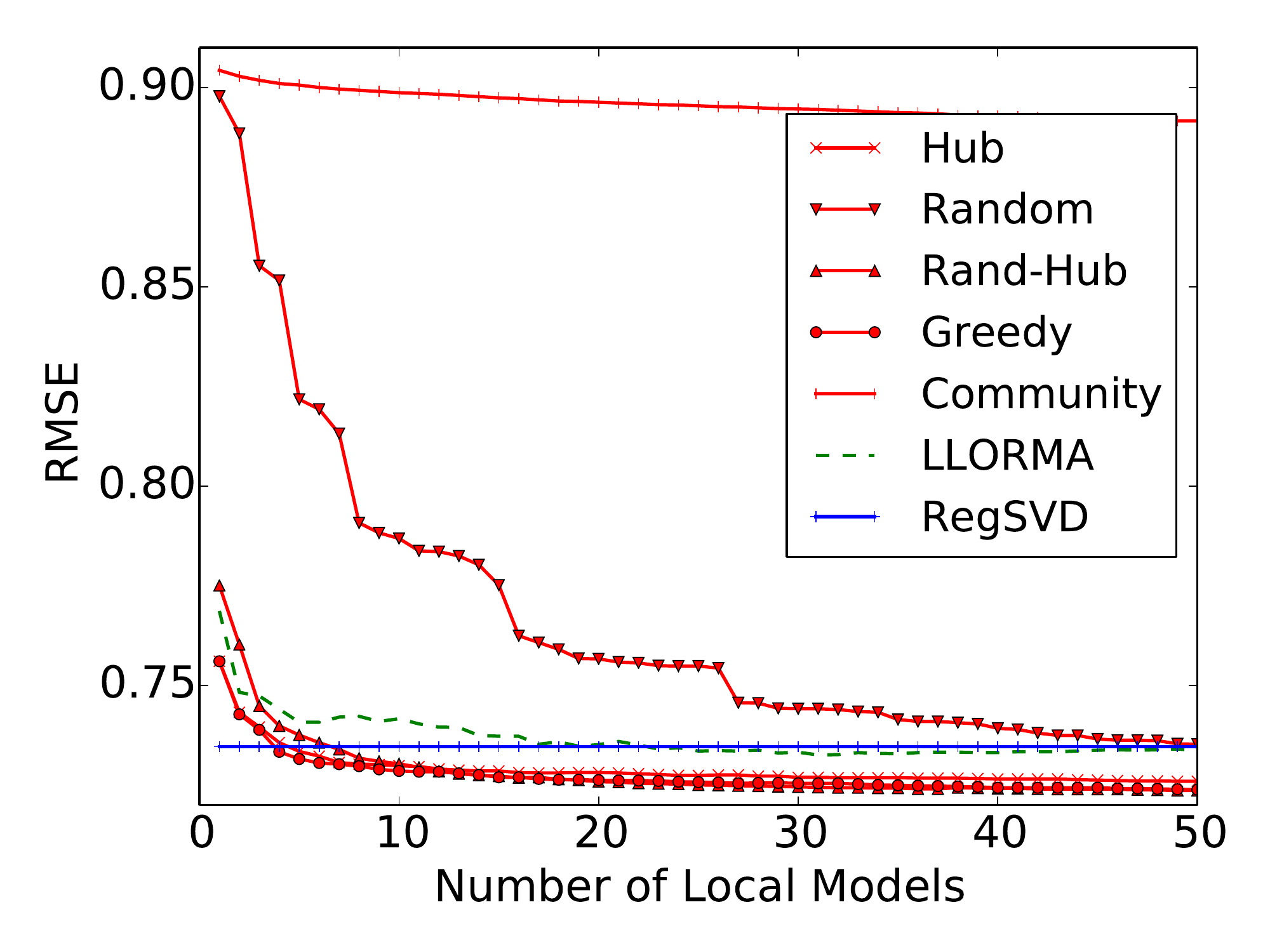}\label{fig-sloma-douban-K10}}
\subfigure[SLOMA($K=20$@Douban).]{\includegraphics[width=0.24\textwidth]{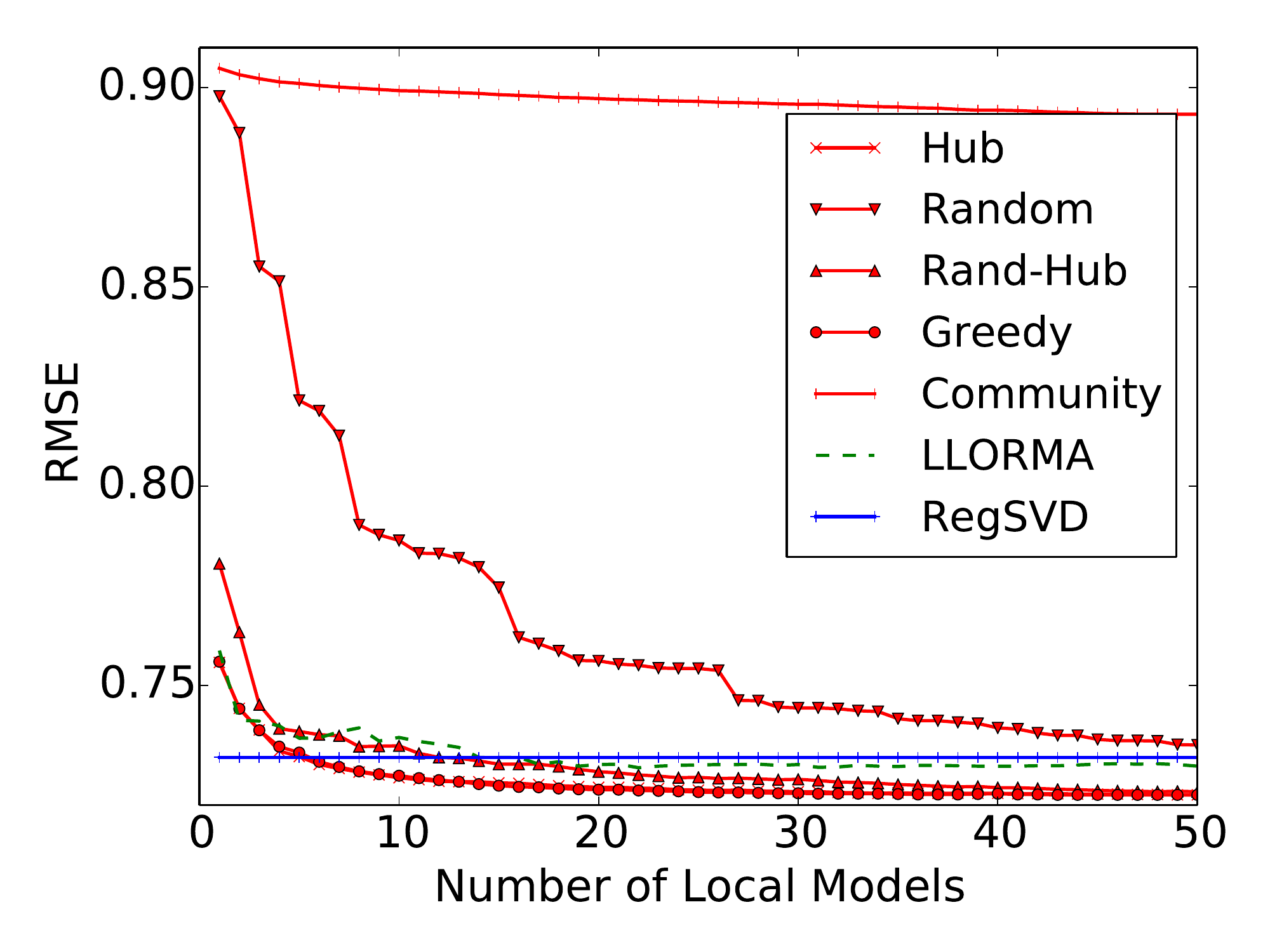}\label{fig-sloma-douban-K20}}
\subfigure[SLOMA++($K=10$@Douban).]{\includegraphics[width=0.24\textwidth]{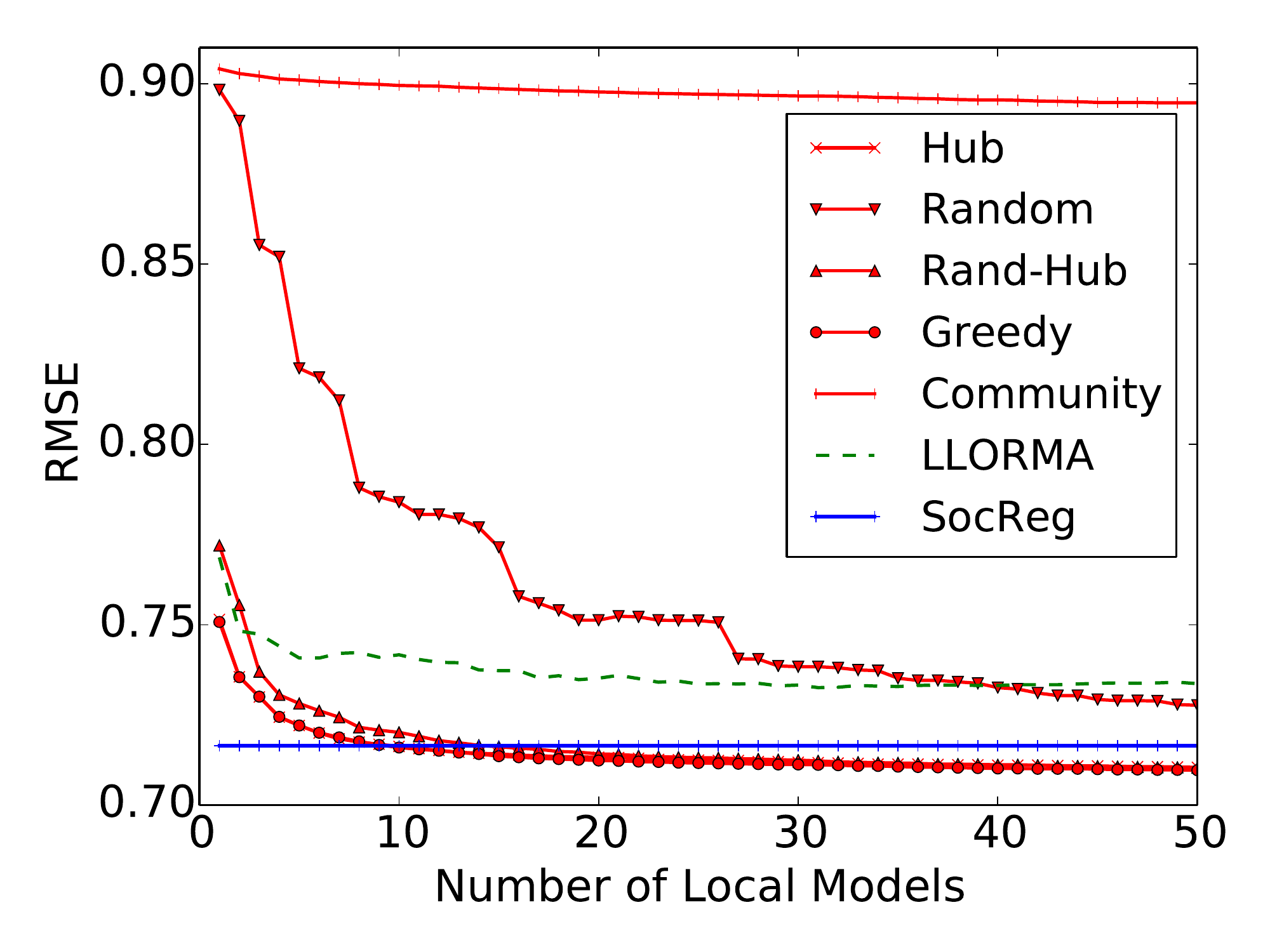}\label{fig-sloma++-douban-K10}}
\subfigure[SLOMA++($K=20$@Douban).]{\includegraphics[width=0.24\textwidth]{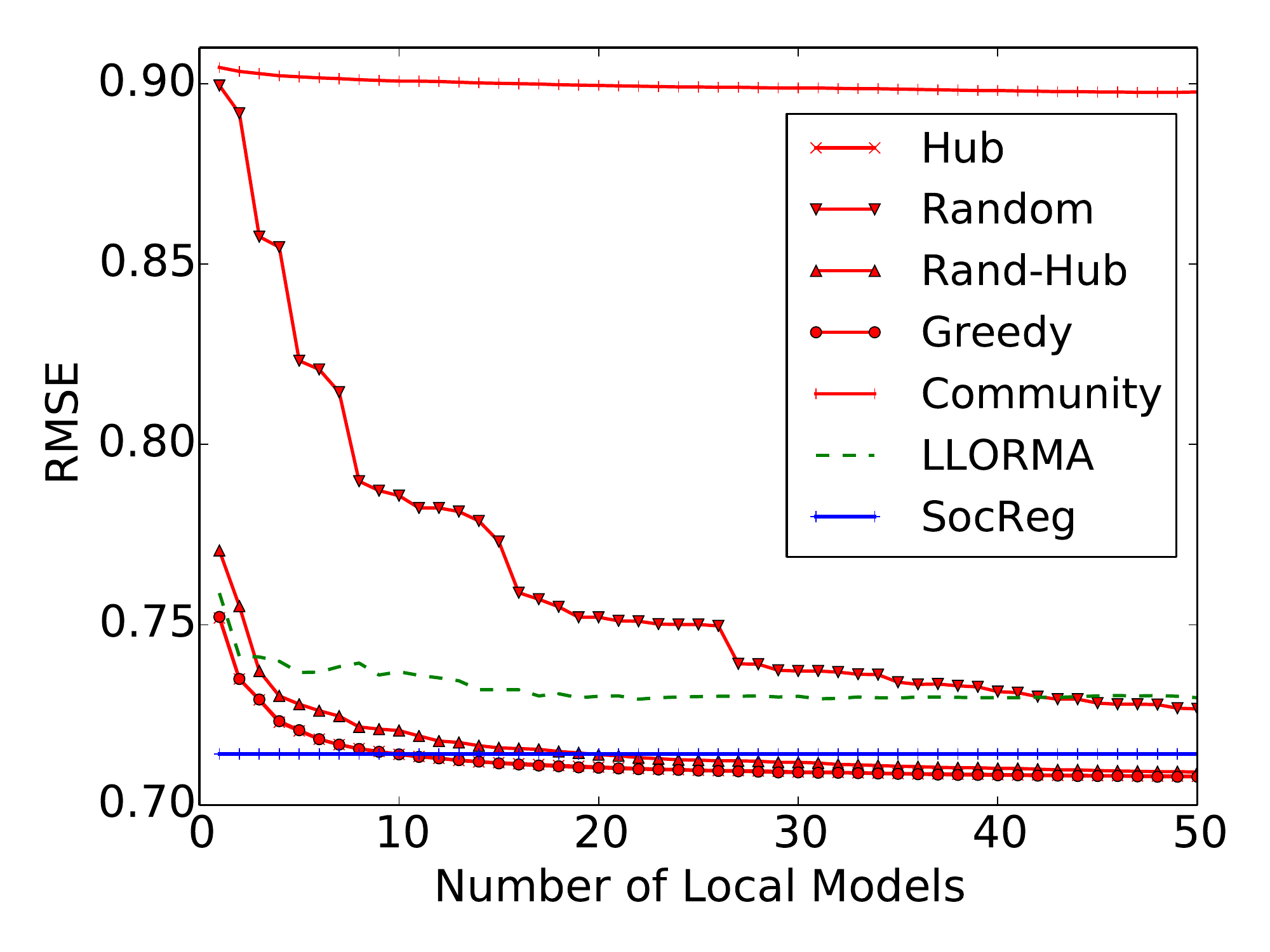}\label{fig-sloma++-douban-K20}}
\caption{RMSEs v.s different methods of identifying connectors on Yelp and Douban.}
\label{fig-vary-connector}
\vspace{-10px}

\end{figure*}

\subsection{Impact of Connector Selection Methods}
\label{subsection-ana-vary-connector}

In this section, we try to answer Question (3) by exploring in depth the performance of different connector selection methods (introduced in Section~\ref{subsec-framework-select-connector}). When adopting the heuristic approaches to construct submatrices, the selection of connectors is very important. It determines the coverages of all local models. As we have elaborated, even though there exist some long-tail users who are rarely selected into any submatrix, we still want to design better heuristic approaches to include as many of them as we can. In Figure~\ref{fig-vary-connector}, we give the comparisons among different selection methods. Note that we also show the results of LLORMA as well as community-based SLOMA, and the results of RegSVD and SocReg are added for SLOMA and SLOMA++, respectively. We set $d = 3$ for different connector selection methods.

From Figure~\ref{fig-vary-connector}, we can see that Hub and Greedy outperform all the
other methods with increasing number of local models. The better performance of Greedy lies
in the fact that we choose uncovered users as connectors with the most neighbors each time, thus we can guarantee that all the connectors will not be too close to each other in the social graph. It will increase the coverage as well as the variance of all the submatrices, which is useful for the ensemble effect of all the local models. However, considering the clarity and simplicity of Hub, we adopt Hub as our selection method in the experiments reported in Section~\ref{subsection-ana-performance}. In Hub, each connector can be regarded as the most active and influential users of the social groups around them, which leads to a low-rank property of the constructed submatrix. Furhter, Hub is simple to implement.

We can see that for all the heuristic approaches, even though Random does not provide good performance, the RMSEs of them decrease with increasing number of local models. The underlying reason is that we can cover more entries in the rating matrix with more local models, leading to the performance gain for overall prediction. Taking all the heuristic approaches and LLORMA into consideration, it further demonstrates the efficacy of local low-rank framework.

\subsection{Discussion on Systematic Submatrices Construction}
\label{subsection-ana-systematic}

From Figure~\ref{fig-vary-connector}, an obvious observation is that the community-based method, i.e., systematic submatrix construction, performs very bad in all settings. We make use of the BIGCLAM~\cite{yang2013overlapping} to detect overlapping communities, each of which corresponds to one submatrix in SLOMA and SLOMA++. The number of communities is set to $50$. However, there are two problems facing the communities detected by BIGCLAM. First, BIGCLAM tends to detect closely connected groups, which leads to smaller coverage of each submatrix and fewer overlaps amongest all the submatrices. As we mentioned above, it impairs the ensemble effects in the overall prediction. Second, there are a small portion of long-tail users (it is around 4\% in our experiments) who cannot be covered by all the communities, thus impairing the performance further. Therefore, it solves Question~(4) as well as explains why we do not report the results of systematic approaches for SLOMA and SLOMA++ in Section~\ref{subsection-ana-performance}. However, we leave it as future work to detect better overlapping communities for SLOMA and SLOMA++.

\section{Conclusions and Future Work}
\label{sec-conclusion}

In this paper, we propose the SLOMA framework, which addresses the problem of submatrix construction facing LLORMA. To the best of our knowledge, it is the first work to incorporate social connections into the local low-rank framework. Based on the social homophily theory, we exploit users' social connections to construct meaningful as well as better submatrices, which leads to superior prediction model compared to LLORMA. Moreover, we integrate social regularization to further strengthen SLOMA. Extensive experiments have been conducted on two real-world datasets, comparing our models against MF as well as LLORMA. The results demonstrate that with the help of social connections, SLOMA outperforms LLORMA and MF.

\section{Acknowledgment}

This project is partially supported by Research Grants Council HKSAR GRF (No.615113) and Microsoft Research Asia. We also thank the anonymous reviewers for their valuable comments and suggestions that help improve the quality of this paper.

\cleardoublepage
\bibliographystyle{IEEEtran}
\bibliography{ref}

\appendices

\section{Proof of Lemma~\ref{lemma-invert}}
\label{lem:proof}

\begin{proof}
Let $P_{\Omega}(A) = A_{ij}$ if $(i,j)\in \Omega$, and $0$ otherwise.
Then,
we can express Equation~\eqref{eq-mf-obj} as
\begin{align}
\min_{\bU,\bV}
\frac{1}{2}\left\| P_{\Omega}(\bU\bV^{\top} - \bO) \right\|_F^2.
\label{eq:temp1}
\end{align}
If $\bU$ and $\bV$ are the optimal solution to Equation~\eqref{eq:temp1},
then they should satisfy first order optimal condition as
\begin{align}
\bU^{\top} P_{\Omega}(\bU\bV^{\top} - \bO) = 0,
\label{eq:temp2} 
\\
P_{\Omega}(\bU\bV^{\top} - \bO) \bV = 0.
\notag
\end{align}
As $\hat{\bU} = \bU \bQ$ and $\hat{\bV} = \bV \bQ^{-1}$,
we have
\begin{align*}
\hat{\bU}^{\top} P_{\Omega}(\hat{\bU} \hat{\bV}^{\top} - \bO) 
& = \bQ^{\top} \bU^{\top} P_{\Omega}(\hat{\bU} \hat{\bV}^{\top} - \bO)
\\
& =  \bQ^{\top} \bU^{\top} P_{\Omega}(\bU \bV^{\top} - \bO) = 0.
\end{align*}
where the last equality comes from Equation~\eqref{eq:temp2}.
Similarly,
we can get
\begin{align*}
P_{\Omega}(\hat{\bU} \hat{\bV}^{\top} - \bO) \hat{\bV} = 0. 
\end{align*}
Thus,
we can see $\hat{\bU}$ and $\hat{\bV}$ also satisfy
the optimal condition of Equation~\eqref{eq-mf-obj},
and then the Lemma holds.
\end{proof}

\end{document}